\documentclass[aps,prd,onecolumn,showpacs,showkeys,groupedaddress,letterpaper]{revtex4}

\usepackage{graphicx}
\usepackage{dcolumn}
\usepackage{bm}
\usepackage{latexsym} 
\usepackage{dsfont}
\usepackage{slashed}
\usepackage{amsmath}

\begin{document}


%
%

\title{Nonleptonic annihilation decays of $B$ meson with a natural infrared
cutoff}

\author{A. A. NATALE and C. M. ZANETTI\footnote{Present address:  
Instituto de F\'{\i}sica, Universidade de S\~ao Paulo, Caixa Postal 66318, 05389-970, S\~ao Paulo, SP, Brazil.} }

\affiliation{Instituto de F\'{\i}sica Te\'orica,
S\~ao Paulo State University, \\
Rua Pamplona, 145 ,
01405-900, S\~ao Paulo - SP -
Brazil\\
email: natale@ift.unesp.br, carina@if.usp.br}


\begin{abstract}
  Within the QCD factorization  approach we compute the amplitudes for
  annihilation  channels   of  B  mesons  decays   into  final  states
  containing two pseudoscalar particles.   These decays may be plagued
  by  effects  like  non-perturbative   physics  or  breaking  of  the
  factorization hypothesis and imply, at some extent, the introduction
  of an infrared  cutoff in the calculation of  amplitudes. We compute
  the decays  with the help  of infrared finite gluon  propagators and
  coupling constants that were  obtained in different solutions of the
  QCD  Schwinger-Dyson  equations.  These  solutions  yield a  natural
  cutoff for the  amplitudes, and we argue that  a systematic study of
  these B decays may provide a test for the QCD infrared behavior.

\end{abstract}


\maketitle

\section{Introduction}

One  of the  main theoretical  challenges  to understand  the B  meson
phenomenology  resides  in  the  matrix elements  calculation  of  the
non-leptonic decay channels.  Several  methods were developed in order
to  deal with  these decays.   The first  to treat  this  problem were
Bauer, Stech  and Wirbel, proposing the  factorization assumption (FA)
and     applying     it     successfully     to     several     decays
\cite{fa}-\cite{fa2}. However,  this method fails  when there are
important  non-factorizable contributions.  Different  approaches were
developed  later  in   order  to  make  progress  in   these  type  of
calculations. One way to proceed, based on the collinear factorization
theorem,  is  the perturbative  QCD  (pQCD)  formalism  as applied  by
Brodsky {\it et al.}  \cite{brodsky},  where it is considered that the
non-leptonic  decays  of heavy  mesons  are  dominated  by hard  gluon
exchange and the  amplitudes are computed based on  the hard exclusive
hadronic   scattering  analysis  developed   by  Lepage   and  Brodsky
\cite{lepage,lepagea}.  A  powerful method  to deal with  these decays
has     been      proposed     by     Beneke      {\it     et     al.}
\cite{beneke}-\cite{beneke3}, which results from the union of the
FA  and pQCD in  the collinear  approximation, hereafter  indicated by
QCDF  (QCD factorization).   These authors  argue that  the transition
form  factors, that  enter in  the  calculation of  $B$ meson  decays,
cannot  be  obtained through  perturbative  methods  because they  are
dominated by ``soft''  interactions, which is a problem  that the pQCD
advocates  claim  to be  solved  by  the  suppression due  to  Sudakov
factors.  On  the other  hand, the non-factorizable  contributions are
dominated by hard gluon exchanges.   Therefore, in QCDF the terms that
are dependent on  the transition form factors are  parameterized as in
the   FA  formalism,  and   the  non-factorizable   contributions  are
perturbatively calculated in the collinear approximation.

Unfortunately,  the formalisms based  on the  collinear approximations
fail   in   the  cases   where   there  are   non-fac\-to\-ri\-za\-ble
contributions related to interactions with the spectator quark and for
annihilation   interactions  due   to  the   appearance   of  endpoint
divergences,  respectively  at  twist-3  and twist-2,  indicating  the
presence  of   non-perturbative  physics  and  the   breaking  of  the
factorization hypothesis \cite{beneke2}.   In the QCDF formalism these
divergences can  be parameterized at  the cost of some  uncertainty in
the calculation.  These contributions are power suppressed in general,
and are  negligible in front of the  factorizable contributions.  This
means that QCDF can be  successfully applied by simply neglecting such
contributions.   However, annihilation  diagrams  can generate  strong
phases that  are relevant for  the CP violation.  Besides  that, there
are decays  that occur only through annihilation  diagrams.  These are
quite  rare  decays,   and  most  of  them  are   beyond  the  present
experimental limits.  It is expected  that they will be detected soon,
with the advent of LHCb (LHC, CERN), or even earlier by the experiment
CDF (Tevatron, Fermilab).   These are the decays that  we expect to be
sensitive to the infrared QCD behavior.

In   this  work   we   compute   some  of   the   decay  channels   of
$B^0_s(\bar{B}^0)$ and  $B^0_d(\bar{B}_d^0)$ which occur  only through
the  annihilation  diagrams: $B_s^0\to  \pi^+\pi^-,\,D^{\pm}\pi^{\mp}$
and  $B_d^0\to   K^+K^-,\,D_s^{\pm}K^{\mp}$.   Working  in   the  hard
scattering collinear approximation of  Brodsky and Lepage, we will use
non-perturbative  solutions  for  the  gluon propagator  and  for  the
running  coupling constant  that were  obtained with  the help  of the
gluonic  Schwinger-Dyson   equations  (SDE).   These  non-perturbative
solutions lead  to a gluon  propagator and coupling constant  that are
finite  in the  infrared, and  this fact  will naturally  supplant, at
leading  twist, the  divergences  that we  discussed  in the  previous
paragraph by providing a natural infrared cutoff.

The  phenomenology  of  $B$  mesons  decays is  known  and  thoroughly
discussed in  Ref.~\cite{buras}.  However, the same  cannot be said
about   the  use   of  SDE   solutions  in   the   strong  interaction
phenomenology, and  for this reason  it is interesting to  recall some
aspects of this  approach.  The SDE form an  infinite tower of coupled
equations that in order to be  solved must be truncated at some order.
Most of the  solutions obtained so far are  gauge dependent and differ
among themselves. It is not surprising that SDE for the QCD propagator
and coupling constant lead to  different solutions as long as they are
solved  with different  truncations  and approximations.   There is  a
class of solutions  for the gluon propagator that  goes to $D(k)=0$ as
the momentum  $k^2 \rightarrow  0$, as well  as another  solution that
goes to  a value different  from $0$ at  the origin of  momenta, whose
inverse value  is usually  denominated as a  dynamical gluon  mass.  A
discussion  about  the  different   solutions  can  be  found  in  the
introduction of  Ref.~\cite{natale}, where the  main references are
also quoted. Only very recently  it was claimed that a gauge invariant
truncation of  the QCD SDE  can be obtained  systematically \cite{bp}.
It is  important to  stress that the  infrared finite behavior  of the
gluon  propagator  has  also  been confirmed  by  lattice  simulations
\cite{lattice,bogolubsky}, with strong evidence for a gluon propagator
with  a   value  at  the   origin  of  momenta  different   from  zero
\cite{cucchieri}.   Accepting  the evidences  for  an infrared  finite
gluon  propagator and  coupling constant,  now we  are faced  with the
problem  of  how to  introduce  these  quantities in  phenomenological
calculations.

There are several  calculations where the SDE solutions  for the gluon
propagator  and  coupling   constant  were  used  in  phenomenological
problems,               see,               for               instance,
Ref.~\cite{natale3}--\cite{brodsky2}.  In all these calculations
the fact that the gluon propagator and coupling constant are finite is
essential for  the result, in such  a way that no  good agreement with
the experimental data is  obtained without appealing to such behavior.
This happens, for  example, in the case of  a non-perturbative Pomeron
model  \cite{natale3,natale3a} or  in a  pion form  factor calculation
\cite{natale4}.  Note  that an infrared finite  coupling constant also
appears in the  context of the so called  Analytic Perturbation Theory
\cite{apt}-\cite{aptc}, and  it improves considerably  the series
convergence of  QCD perturbative calculations.  On the  other hand, we
also  expect that  any infrared  finite  gluon propagator  leads to  a
freezing  of the  infrared coupling  constant  \cite{natale2}, meaning
that  the  use  of  an   infrared  finite  gluon  propagator  must  be
accompanied  by an  infrared finite  coupling constant.   These finite
expressions for  the coupling constant and gluon  propagator should be
used in actual calculations in the sense of the Dynamical Perturbation
Theory   (DPT)  proposed  by   Pagels  and   Stokar  many   years  ago
\cite{pagels}.   A  finite coupling  can  also  be  interpreted as  an
effective   charge,   as  performed   in   several  perturbative   QCD
calculations  \cite{brodsky3,brodsky3a}.  Finally,  the characteristic
mass scale  of these solutions  is at least a  factor $\mathcal{O}(2)$
above the QCD scale ($\Lambda_{{QCD}}$), and for some of the solutions
the freezing  of the coupling  constant happens at a  relatively small
value     \cite{natale4,cornwall2007}.     According     to    Brodsky
\cite{brodsky4,brodsky4a},  the  inclusion  of  such effects,  as  the
freezing of the QCD running coupling at low energy scales, could allow
to reliably  capture the non-perturbative QCD effects  at an inclusive
level.

The physics of  non-leptonic decays of $B$ mesons  may furnish another
important  test in the  quest of  learning about  the non-perturbative
behavior of  the gluon propagator  and the running  coupling constant.
One   of   these   non-perturbative   SDE  solutions   (the   one   of
Ref.~\cite{cornwall}) has  already been  used to compute  the decay
amplitudes of  $B$ mesons \cite{yang}-\cite{yangb},  although the
effect  of the associated  infrared finite  coupling constant  was not
taken into  account in  these studies.  In  this work we  test several
solutions and  show that  they lead to  different predictions  for the
branching  ratios of  $B$ meson  decays.   Of course,  there are  many
physical quantities that can also  modify the decay rates, but keeping
unchanged  all the same  quantities (like  wave functions  and others)
while  varying only the  coupling constants  and gluon  propagators we
come to  the conclusion that a  systematic study of  these decays will
help to select a specific QCD infrared behavior.
 
This paper is organized as following: In Sec.~2 we pre\-sent the basic
approach to  study the  non-leptonic $B$ meson  decays.  In  Sec.~3 we
show  the annihilation  amplitudes for  the different  $B$  decays. In
Sec.~4  we discuss  the different  non-perturbative solutions  for the
gluon  propagator  and   the  running  coupling  constant.   Section~5
contains  our  numerical results  and  Section  6  is devoted  to  our
conclusions.


\section{Hamiltonian and amplitudes for $B$ decays}

Weak decays of $B$ mesons are described by an effective Hamiltonian at
a renormalization scale $\mu\ll M_W$.  For a $B$ meson decaying into a
final state $f$ the effective Hamiltonian is given by \cite{buras}:
\begin{equation}\label{ham}
{\mathcal H}_{\mathrm{eff}}=\frac{G_F}{\sqrt{2}}V_{\mathrm{CKM}}\sum_iC_i(\mu)\,Q_i(\mu),
\end{equation}
where   $G_F$   is   the   Fermi  constant,   $V_{\mathrm{CKM}}$   are
Ca\-bib\-bo-Ko\-ba\-ya\-shi-Mas\-ka\-wa   factors,   $Q_i$   are   the
operators  contributing to  the decay  and $C_i(\mu)$  are  the Wilson
coef\-fi\-ci\-ents.

The operators  depends on the flavor  structure of the  decay. In this
work we will discuss two types of $B$ meson decays: the charmed decays
characterized  by transitions  with $\Delta  B=1$, $\Delta C=\pm  1$, and
charmless decays characterized by $\Delta B=1$, $\Delta C=0$.

For   charmed   decays,   there  are   only  contributions   from
current-current operators:
\begin{eqnarray}\label{qcharmed}
Q_1  =  (\bar{b}_ic_j)_{\mathrm{V-A}}(\bar{u}_ir_j)_{\mathrm{V-A}},\,\,\,
Q_2  =  (\bar{b}_ic_i)_{\mathrm{V-A}}(\bar{u}_jr_j)_{\mathrm{V-A}}
\end{eqnarray}
For charmless decays, the operators can be divided as:

\noindent {\it 1. Current-current operators:}
\begin{eqnarray}\label{qcharmless}
Q_1 = (\bar{r}_iu_j)_{\mathrm{V-A}}(\bar{u}_jb_i)_{\mathrm{V-A}},\,\,\,
Q_2 = (\bar{r}_iu_i)_{\mathrm{V-A}}(\bar{u}_jb_j)_{\mathrm{V-A}}.
\end{eqnarray}
\noindent {\it 2. QCD penguin operators:}
\begin{eqnarray}\label{pqcdcharmless}
Q_3  =  (\bar{r}_ib_i)_{\mathrm{V-A}}\sum_{q}(\bar{q}_jq_j)_{\mathrm{V-A}}\,;\\
Q_4  =  (\bar{r}_ib_j)_{\mathrm{V-A}}\sum_{q}(\bar{q}_jq_i)_{\mathrm{V-A}}\,;\nonumber\\
Q_5  =  (\bar{r}_ib_i)_{\mathrm{V-A}}\sum_{q}(\bar{q}_jq_j)_{\mathrm{V+A}}\,;\nonumber\\
Q_6  =  (\bar{r}_ib_j)_{\mathrm{V-A}}\sum_{q}(\bar{q}_jq_i)_{\mathrm{V+A}}\,.\nonumber
\end{eqnarray}
\noindent{\it 3. Electroweak penguin operators:}
\begin{eqnarray}\label{pewcharmless}
Q_7 = \frac{3}{2}(\bar{r}_ib_i)_{V-A}\sum_{q}e_{q}(\bar{q}_jq_j)_{V+A}\,;\\
Q_8 = \frac{3}{2}(\bar{r}_ib_j)_{V-A}\sum_{q}e_{q}(\bar{q}_jq_i)_{V+A}\,;\nonumber\\
Q_9 = \frac{3}{2}(\bar{r}_ib_i)_{V-A}\sum_{q}e_{q}(\bar{q}_jq_j)_{V-A}\,;\nonumber\\
Q_{10} = \frac{3}{2}(\bar{r}_ib_j)_{V-A}\sum_{q}e_{q}(\bar{q}_jq_i)_{V-A}\,,\nonumber
\end{eqnarray}	
\noindent  where $i,j$  are  color indices,  $r=d$ or $s$,  $e_q$ is  the
electric  charge  in $\vert  e\vert$  units and  $(\bar{q_1}q_2)_{V\pm
A}=\bar{q}_1\gamma_{\mu}(1\pm\gamma_5)q_2$.     The    current-current
operators describe the  $W$ boson exchange at tree  level, whereas the
penguin  operators  occur at  the  1-loop  level  and describe  flavor
changing neutral currents.

The decay amplitude for a  $B$ meson into a final state $f$ is obtained
from (\ref{ham})
\begin{eqnarray}\label{amp}
{\mathcal  A}(B\to f) & = & \langle  f\vert{\mathcal
H}_{\mathrm{eff}}\vert B\rangle \nonumber\\
&= & \frac{G_F}{\sqrt{2}}\sum_i V_{\mathrm{CKM}}^i C_i(\mu)\langle f \vert Q_{i}\vert B\rangle(\mu),
\end{eqnarray}
\noindent  where  $\langle  f\vert   Q_{i}  \vert  B\rangle$  are  the
ha\-dro\-nic ma\-trix  e\-le\-ments be\-twe\-en the  initial and final
states.  Due  to the existence  of long distance  interactions between
hadrons   in  non-leptonic  decays,   the  matrix   elements  $\langle
Q_i\rangle$  must   be  obtained  through   factorization  schemes  or
non-\-per\-tur\-ba\-ti\-ve    cal\-cu\-la\-tions,   like   lat\-ti\-ce
si\-mu\-la\-tion,  QCD  sum rules,  chiral  perturbation theory,  {\it
  etc}.  The literature contains  a long discussion about these matrix
elements  and  their  contributions  to  the  different  factorization
methods that  we quoted before (FA,  QCDF and pQCD).  

In the  collinear approximation worked  out by Brodsky and  Lepage the
matrix elements of Eq. (\ref{amp}) are obtained through a convolutions
of the hard  scattering kernel and the distribution  amplitudes of the
mesons involved in the process.  In the case of a two-body nonleptonic
annihilation decay  $B$ $\to$ $M_1M_2$,  where the final state  can be
two light mesons  or a heavy and a light meson,  the matrix element of
an operator $Q_i$ is given by:
\begin{eqnarray}\label{collinear}
\langle  f \vert  Q_i\vert   B\rangle  =   \int_0^1d[\chi]\,T_i([\chi])\,\Phi_{\mathrm{M}_1}(x)\,\Phi_{\mathrm{M}_2}(y)\,\Phi_{\mathrm{B}}(z)\,,
\end{eqnarray}
\noindent   where   $[\chi]=x,y,z$   are   the   momentum   fractions,
$\Phi_{\mathrm{M}}(x)$ are the  light-cone distribution amplitudes for
the quark-antiquark sta\-tes of the mesons, which are non-perturbative
functions of  the momentum fraction  carried by the partons;  $T_i$ is
the  hard scattering  kernel that  can be  perturbatively  computed as
function of the light-cone momenta of collinear partons.

The Eq.   (\ref{collinear}) is used  in the QCD  factorization approach
\cite{beneke}-\cite{beneke3},  introduced  by  Beneke,  Buchalla,
Neubert   and  Sa\-chraj\-da   (BBNS)   to  compute   non-factorizable
contributions  such as hard  spectator interactions  and annihilation.
In  order  to  have  an actual  factorization,  the  di\-ver\-gen\-ces
originated  in  the  non-factorizable  con\-tri\-bu\-tions  should  be
absorbed by  the distribution amplitudes. However,  there are endpoint
divergences  that  break   down  the  factorization.   These  endpoint
divergences  are  certain  to  appear  in the  kernel  calculation  of
annihilation  contributions.  The  introduction  of a  cutoff is  then
necessary,  and  this  is  performed  by BBNS  through  the  following
parameterization
\begin{equation}
\int    \frac{d   x}{x} = {\mathrm{ln}}\frac{m_B}{\Lambda_h}(1+\varrho
e^{i\varphi}),\,\,\,\,\,\,0\leq\varrho\leq 1.
\end{equation}
\noindent  These  are   the  cutoffs  that  we  referred   to  in  the
introduction. This is an inconsistency of the method that introduces a
large uncertainty  in the  calculation, in such  a way that  we obtain
only an estimate of the amplitudes.

The  end-point singularities that  we discussed  above also  appear in
perturbative   QCD   calculations  of   $B$   meson  transition   form
factors. This type of infrared problem  can be handled in the same way
as will be  discussed in the next sections.   This problem was already
studied by some of us  in the cases of the $\gamma\to\pi_0$ transition
and  $\pi$  meson  form  factors  \cite{natale5a,natale4}.   The  only
difference in the $B$ meson case  is that the effect of dressed gluons
will not be  as pronounced as it  is for the $\pi$ meson  case, due to
the large mass of the $B$ meson that now appears in the calculation.

\section{Amplitudes for non-leptonic annihilation decays}

\begin{figure}[pb]
\includegraphics[width=10cm]{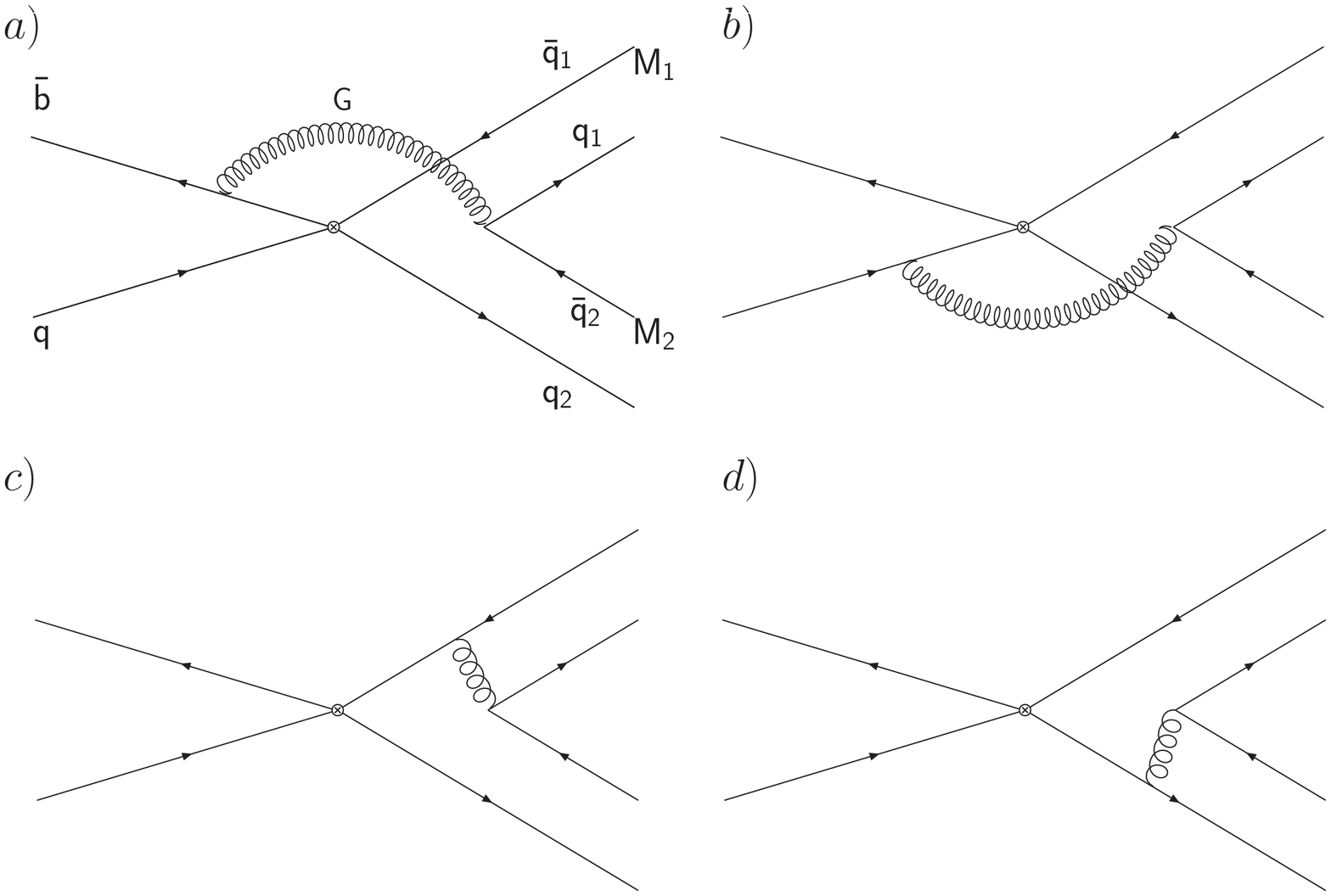}
\vspace*{8pt}
\caption{\label{feynm} Feynman  diagrams contributing to  the amplitude
  of annihilation decay channels.}
\end{figure}

In  this  work  we  will   compute  non-leptonic  B  decays  into  two
pseudoscalar  mesons occurring only  through annihilation  diagrams. A
typical  decay is shown  in Fig.~\ref{feynm},  where the  bottom quark
decays  via $W$ exchange  or penguin  processes, and  a gluon  that is
emitted from any  one of the quarks involved in  the process creates a
quark-antiquark pair in the final state, which will be a leading order
process  in  $\alpha_s$.  The  channels  we will  analyze  with  these
characteristics are:
\begin{itemize}
\item Charmless channels :  $\bar{B}^0_s\to\pi^+\pi^-$;  $\bar{B}^0_d\to K^+K^-$;

\item Charmed channels: $\bar{B}_d^0 \rightarrow  D_s^{\pm}  K^{\mp}$; $\bar{B}_s^0 \rightarrow  D_{}^{\pm} \pi^{\mp}$.
\end{itemize}

To compute the annihilation amplitudes we use the QCDF approach, where
the matrix elements are calculated through the Eq.  (\ref{collinear}).
For  the distribution  amplitudes we  adopt the  expressions  found on
Ref.~ \cite{beneke2}  at leading twist order.   The decay amplitude
$\mathcal{A}  (B\to M_1M_2)$  is  obtained from  Eq.  (\ref{amp})  and
(\ref{collinear}), and the branching ratio is then calculated from
\begin{equation}
{\mathcal B}(B\to M_1M_2)=\frac{\tau_B p_c}{8\pi m_B^2} 
\vert\mathcal{A} (B\to M_1M_2)\vert^2,
\end{equation}
where $\tau_B$ is the $B$ meson  lifetime and $p_c$ is the momentum of
the final state  particles with masses $m_1$ and $m_2$  in the $B$ meson
rest frame, given by
\begin{equation}
p_c=\frac{\sqrt{(m_B^2-(m_1+m_2)^2)(m_B^2-(m_1-m_2)^2))}}{2m_B}.
\end{equation}

As mentioned  before, these amplitudes contain  divergences when dealt
with in  the collinear  approximation, as in  the QCDF  approach. Such
divergences can be eliminated in  the pQCD method when the transverse
momenta     of    the     partons    are     taken     into    account
\cite{pqcd}-\cite{pqcd3}.  However, in this work we will make use
of non-perturbative gluon propagators,  obtained with the help of SDE.
These  gluon propagators  are  finite  in the  infrared,  and so  they
provide  a natural  cutoff to  the  singularities that  appear in  the
collinear approximation.   This kind  of calculation has  already been
performed by  Yang {\it  et al.}  \cite{yang}-\cite{yangb}  for a
specific SDE  solution, and  our intention is  to make a  more general
discussion including  other solutions, and verify how  the results are
dependent  on  the  particular  choices  of SDE  solutions.   We  also
introduce  the  effect  of   the  infrared  finite  coupling  constant
associated to each SDE solution,  which was not considered in previous
works.

\subsection{Charmless channels}

\begin{figure}[pt]
\includegraphics[width=10cm]{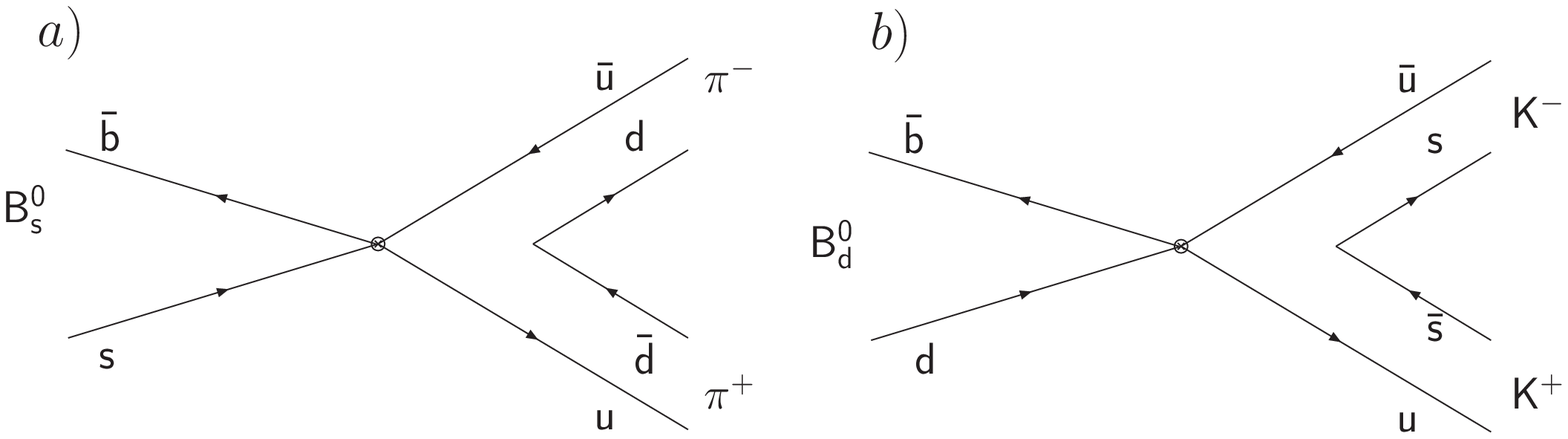}
\vspace*{8pt}
\caption{\label{fig2}  Quark  contents   of  the  decay  channels:  a)
  $B_s^0\to\pi^+\pi^-$,  b) $B_d^0\to  K^+K^-$.}
\end{figure}

There are  two charmless decay  channels whose final states  are light
pseu\-dos\-ca\-lar mesons that occur through pure an\-ni\-hi\-la\-tion
diagrams: $B_s^0$ $\to$ $\pi^+$ $\pi^-$ and $B_d^0$ $\to$ $K^+$ $K^-$.
These  channels happen through  the elementary  processes $b\bar{s}\to
u\bar{u}$ and  $b\bar{d}\to u\bar{u}$, respectively,  and the creation
of a $d\bar{d}$ pair by a  gluon, as is shown in the Figs.~\ref{fig2}a
and  \ref{fig2}b.  These  decays  receive contributions  from the  $W$
exchange operator $Q_2$ and  the penguin operators $Q_4$, $Q_6$, $Q_8$
and         $Q_{10}$,         given         in        the         Eqs.
(\ref{qcharmless})-(\ref{pewcharmless}), with $r=d$  and $r=s$ for the
decays of the $B_d^0$  and $B_s^0$ mesons, respectively. The operators
$Q_2$, $Q_4$  and $Q_{10}$ have a  $(V-A)\otimes(V-A)$ structure while
the operators $Q_6$ and $Q_8$ are of the type $(V-A)\otimes(V+A)$.  

We  compute the  contribution of  each  operator for  the diagrams  of
Fig.~\ref{feynm} with the  help of Eq.(\ref{collinear}).  The diagrams
that contribute for the  amplitude are shown in Figs.~\ref{feynm}a and
\ref{feynm}b,    since   the    contributions   from    the   diagrams
Figs.~\ref{feynm}c  and \ref{feynm}d cancel  among themselves.   The
full amplitude for these decays is given by
\begin{eqnarray}\label{ampcharmless}
{\mathcal A}(B\to f)& = &\frac{G_F}{\sqrt{2}}f_Bf_{M_1}f_{M_2}\pi\frac{C_F}{N_C^2}\,( V_{ub}\,V_{ur}^*\,{\mathcal A}_{\mathrm{Tree}} - V_{tb}\,V_{tr}^*\,{\mathcal A}_{\mathrm{Penguin}}),\nonumber
\end{eqnarray}
\noindent with  $r=d$ and $r=s$ for  the decays of  the $B_d^0$
and $B_s^0$ mesons, respectively.  The tree level and penguin amplitudes are
\begin{equation}
{\mathcal A}_{\mathrm{Tree}}=C_2\,A_{1}
\end{equation}
\noindent and 
\begin{equation}
{\mathcal A}_{\mathrm{Penguin}}=\left( 2\,C_4 + \frac{C_{10}}{2}\right)\,A_{1}+\left(2\,C_6 + \frac{C_8}{2}\right)\,A_{2}.
\end{equation}
The functions $A_1$ and $A_2$  are the contributions from operators of
the  type $(V-A)\otimes(V-A)$  and  $(V-A)\otimes(V+A)$, respectively,
and are given by
\begin{eqnarray}\label{A1}
A_{1}& = &  \int_0^1 d[X]\,\phi_{M_1} (x) \,\phi_{M_2} (y)\, \phi_B (z) \,D_g(x,y) \bigl[ \alpha_s(\mu_h)(y-z) D_q(x,y,z) + \nonumber\\ & &+ \alpha_s(\mu)(1-x)\,D_b(x,y,z) \bigr] ,
\end{eqnarray}
\begin{eqnarray}\label{A2}
A_{2} &= & \int_0^1  d[X],\phi_{M_1} (x)\,\phi_{M_2} (y)  \phi_{B} (z)\,D_g(x,y) \bigl[  \alpha_s(\mu_h)\,(\bar{x}-z) \,D_q(x,y,z) + \nonumber\\ && + \alpha_s(\mu)\,y \,D_b(x,y,z) \bigr]\,,
\end{eqnarray}

\noindent where $d[X]=dx\,dy\,dz$, the quark propagators $D_b$ and
$D_q$ ($q=d$ and $s$ for the $B_d^0$ and $B_s^0$ mesons, respectively)
are given by:
\begin{eqnarray*}
D_b^{-1}(x,y,z)&=&(1-(x-z)(\bar{z}-y));\\
D_q^{-1}(x,y,z)&=&(\bar{x}-z)(y-z),
\end{eqnarray*}
\noindent and $D_g(x,y)$ is the gluon propagator, whose the perturbative expression is $D_g^{-1}=k_g^2= y\,(1-x)$.

The  scales  adopted  in   the  calculations  are  $\mu=m_b$  for  the
contribution  from the  diagram shown  in Fig.~\ref{feynm}a  where the
gluon is  emitted from the $b$  quark, and $\mu_h=\sqrt{\Lambda_hm_b}$
for  the contribution shown  in Fig.~\ref{feynm}b  where the  gluon is
emitted  from  the   ``spectator''  quark,  with  $\Lambda_h=500$  MeV
\cite{beneke2}.    When   the  effective   scale   $\mu<m_b$  we   use
$\Lambda_{\mathrm{QCD}}  =  225$ MeV  and  when  $\mu  = m_b$  we  use
$\Lambda_{\mathrm{QCD}} = 300$ MeV, and the difference is to match the
values of the coupling constant due to different quarks thresholds. At
the Eq.~(\ref{ampcharmless}), the scales of the Wilson coefficients
for each term are equal to the ones of the coupling constants.

\subsection{Charmed channels}

\begin{figure}[pt]
\includegraphics[width=10cm]{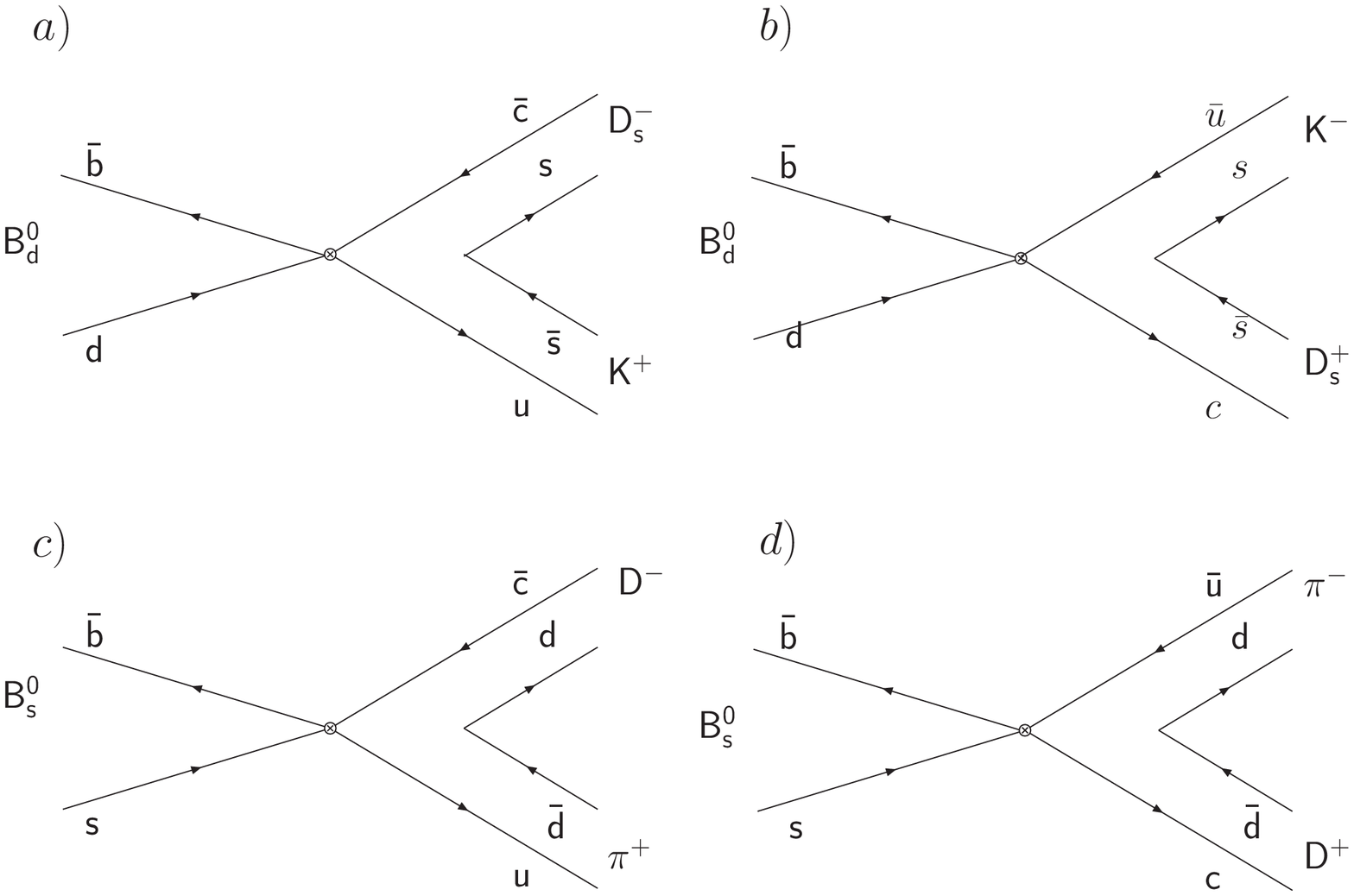}
\vspace*{8pt}
  \caption{\label{figura3} Quarks contents of the decay channels:
    a) $B_d^0$ $\rightarrow  D_s^-K^+$, b) $B_d^0 \rightarrow D_s^+K^-$,
    c) $B_s^0  \rightarrow D^-\pi^+$, d)  $B_s^0 \rightarrow D^+\pi^-$}
\end{figure}

The charmed  decay channels occur through $W$  exchange processes, and
there is  no contribution from  penguin diagrams.  The  operators that
contribute   for  the   effective   Hamiltonian  are   given  by   the
current-current operators $Q_1$  and $Q_2$ (Eq.~(\ref{qcharmed})). The
Fig.~\ref{figura3}a  and \ref{figura3}b shows  the quark  diagrams for
the  decay channels  $B_d^0  \rightarrow D_s^\mp  K^\pm$, which  occur
through    the   elementary   processes    $\bar{b}d\to\bar{c}u$   (or
$\bar{b}u\to\bar{c}d$).   The  quark diagrams  for  the decay  channel
$B_s^0 \rightarrow  D^\mp\pi^\pm$ are shown  in the Fig.~\ref{figura3}c
and   \ref{figura3}d,   occurring   though   the   elementary   process
$\bar{b}s\to\bar{u}c$.

The full  amplitude for the general  processes $B^0\to D^{\mp}M^{\pm}$
(with  $B=B_d$ and $B_s$,  $D,M={D_s,K},{D,\pi}$) will  be calculated
from
\begin{eqnarray}                  {\mathcal                  A}(B^0\to
  D^{\mp}M^{\pm})=\frac{G_F}{\sqrt{2}}f_Bf_{D}f_{M}\pi\frac{C_F}{N_C^2}
  V_{CKM}A_{B^0\to D^\mp M^\pm}\,,
\end{eqnarray}
The     CKM     factors     for     each    decay     channel     are:
$V_{\mathrm{CKM}}=V_{cb}^*V_{ud}$   for    the   $D_s^-K^+$   channel;
$V_{ub}^*V_{cd}$ for  $D_s^+K^-$; $V_{us}V_{cb}^*$ for  $D^-\pi^+$ and
$V_{cs}V^*_{ub}$  for $D^+\pi^-$.   The functions  $A_{{B^0  \to D^\mp
    M^\pm}}$ are given by the following integrals:
\begin{eqnarray}\label{BDK1}
 A_{{B^0 \to D^-M^+}} &  =  & \int_0^1   {dx}\,{dy}\,{dz} \,\phi_{D} (y) \, \phi_M (x)\,D_g(x,y)
\,\biggr\{  \alpha_s(\mu_f) \biggl(   C_1(\mu_f)  +   \frac{C_2(\mu_f)}{2}   \biggl)\times \nonumber\\ 
&\times&[ - (1-r^2)\bigl( (1-2r^2-(1-r^2)\,x) \,  D_c(x)-y\, D_u(y) \bigr) \bigr]+ \phi_B(z)\times \nonumber\\ 
&\times&\bigl[
  \alpha_s(\mu_h)\, C_2(\mu_h)\,(1-r^2)\, \bigl( (1-r^2)
\,(1-x)+r^2\,(y-z) \bigr)\,D_d(x,y,z)+\nonumber\\
&-&\alpha_s(\mu)\,C_2(\mu)(1-r^2)\, \bigl( (1+r^2)\,(y+z) -r^2 \bigr)\,D_b(x,y,z)\bigr]\biggr\}\,,
\end{eqnarray}
and
\begin{eqnarray}\label{BDK2}
A_{B_d^0\to D^+M^-} &  = & \int_0^1  {dx}\,{dy}\,{dz} \,\phi_{D} (y) \, \phi_M (x)\,D_g(x,y)\biggr\{    \alpha_s(\mu_f) \biggl(   C_1(\mu_f)   +   \frac{C_2(\mu_f)}{2}   \biggl)\times\nonumber\\
&\times&[   (1-r^2)\bigl( (1-2r^2-(1-r^2)\,x) \,  D_c(x)-  \,y\, D_u(y) \bigr)  \bigr]+ \Phi_B(z)\times+\nonumber\\
&+& \bigl[\alpha_s(\mu_h) C_2(\mu_h) ( 1-r^4)\,(y-z)\, D_d(x,y,z)-\alpha_s(\mu)C_2(\mu)(1-r^2)\times\nonumber\\
&\times& \bigl( (1-r^2)\,(1-x)-r^2+r^2\,(y+z) \bigr)\,D_b(x,y,z)\bigr]\biggr\}\,.
\end{eqnarray}

\noindent The functions $D_b(x,y,z),\,D_d(x,y,z),\,D_u(y)$
and $D_c=D_c(x)$ are the quarks propagators, given by:
\begin{eqnarray*}
 D_b^{-1}(x,y,z)&=& y+z+(1-y-z)(1-x)(1-r^2); \\ D_d^{-1}(x,y,z)& = & x(z-y)(1-r^2);\\
 D_c^{-1}(x)&=& (1-x)(1-r^2)\,;\\ D_u^{-1}(y)& = & y(1-r^2)\,.
\end{eqnarray*}
\noindent  The  function $D_g(x,y)$  is  the  gluon propagator,  whose
perturbative  expression  is  $D_g^{-1}=k_g^2 =  y\,(1-x)\,(1-r^2)$.
Note  that the integrals  in the  former equations  (and also  in Eqs.
(\ref{A1}) and (\ref{A2}))  are dimensionless after the simplification
of  the $m_B$  dependence in  both numerator  and  denominator.

The scales adopted  in the calculations are the  same as the charmless
case: $\mu=m_b$ for the contribution of the diagram where the gluon is
emitted from the $b$ quark,  and $\mu_h = \sqrt{\Lambda_hm_b}$ for the
contribution where the gluon is emitted from the ``spectator'' quark, with
$\Lambda_h=500$  MeV.   For  the  diagrams  of  Fig.~\ref{feynm}c  and
\ref{feynm}d where the  gluon is emitted from the  quarks in the final
state, we use  the scale $\mu_f=\mu/2=m_b/2$. The QCD  scales are also
the same as the charmless case.

\section{Non-perturbative gluon propagator and coupling constant}

To compute  the amplitudes discussed in the  previous section, instead
of  the   perturbative  expression  $D_g=(k^2)^{-1}$   which  is  IR
divergent,  we  will use  for  the  gluon  propagator $D_g(x,y)$  that
appears   in  Equations   (\ref{A1}),  (\ref{A2}),   (\ref{BDK1})  and
(\ref{BDK2}) different  solutions of the  gluonic SDE obtained  in the
literature.   Each one  of these  solutions is  also associated  to an
infrared  finite coupling  constant  that enters  in the  calculation.
This coupling constant  is evaluated at the scale  $\mu$, which is the
average energy of the gluon emission,  where we may still have a small
difference  between the different  solutions for  the non-perturbative
couplings.   The propagators  have the  interesting property  of being
infrared  finite eliminating  the  endpoint divergences,  and at  high
energies they  match with their perturbative  counterpart.  Note that,
as discussed  in Ref.~\cite{natale},  in these calculations  we are
integrating over the  gluon propagator in a large  region of the phase
space,  and this  integration  is weighted  by different  distribution
functions,  therefore  the result  will  depend  non-trivially on  the
formal  expression of the  propagator, as  well as  it will  depend on
different values of the non-perturbative coupling constant.  Even with
the large mass  scale involved in the problem (the  $B$ meson mass) we
may expect a signal of the infrared behavior of these quantities.

Notice that the behavior of  the coupling constant and the propagators
are  intimately connected.   As shown  by Cornall  \cite{cornwall} the
product $g^2 D(q^2)$  is constant, in a more  detailed discussion this
can also be  verified in the recent work  by Aguilar and Papavassiliou
\cite{aguilar2a,aguilar2b,aguilar08}. Furthermore Alkofer  et al. in a
series  of papers  \cite{alkofer}-\cite{alkoferc}  have advocated
that $\alpha_s (q^2)=(g^2/4\pi)Z_D(q^2) Z^2_G(q^2)$  (the Z' s are the
gluon  and ghost  propagators  dressing).  Such  type  of relation  is
expected in the dynamical mass  generation mechanism and can be traced
back  to the  Slavnov-Taylor or  Ward-type identities.   Therefore the
determination of the actual  infrared behavior imply in the associated
effect of the coupling constant and propagator!

It  could be  asked if  there is  double counting  of non-perturbative
dynamics  between QCDF  and the  employment of  infrared  finite gluon
propagators,   since   according   to   the   factorization   theorem,
non-perturbative dynamics in B meson decay processes has been absorbed
into distribution  amplitudes.  We stress  that this is not  the case.
The effect of  the infrared finite gluon propagators  affects only the
hard part  of the  hadronic matrix elements.   The introduction  of an
infrared finite gluon propagator means  that as we lower the energy we
cannot neglect  anymore the  gluon dressing that  is predicted  by the
SDE.   Therefore dressed  gluons provide  the natural  cutoff  for the
theory,  and should be  considered in  the calculation  as anticipated
many years ago  in the proposal of a  dynamical perturbation theory by
Pagels  and Stokar  \cite{pagels}.  Only  the hard  scattering kernels
(like  $T^I$  and  $T^{II}$  in  Eq.(6)) will  be  affected,  and  the
distribution   amplitudes  should   naturally   contain  the   typical
non-perturbative  dynamics  of  the  processes involving  meson  bound
states.

The usual  treatment of B meson decays  involves consistent expansions
in $\alpha_s$ and in $1/m_b$.   Such level of consistency has not been
achieved yet in the case of solutions of SDE for the gluon and fermion
propagators. First, only quite  recently it was developed a systematic
gauge invariant  approximation and  truncation method for  the gluonic
Schwinger-Dyson  equation  (see  the  work of  Ref.~\cite{bp}).   This
method is consistent with  an $\alpha_s$ expansion.  Secondly, to also
have  an expansion  consistent with  a $1/m_b$  expansion for  B meson
decays  it will  be  necessary to  solve  the coupled  Schwinger-Dyson
equations for  the gluon  and fermion propagators,  and this  is quite
away from the present status of the theory. The only point that can be
guessed at  the moment is that  the results for  infrared finite gluon
propagator will be  only slightly changed by fermionic  effects as far
as the number of quarks remains equal to the known one.

In  this  section we  present  the  different  SDE solutions  for  the
coupling constant and gluon  propagator $(D(q^2))$. They were obtained
as the  result of different approximations  made to solve  the SDE for
pure gauge  QCD. There are two  classes of solutions  according to the
behavior of  the gluon  propagator at the  origin of the  momenta, and
both lead  to an  infrared frozen coupling  constant: (a)  an infrared
finite gluon propagator which is  different from zero at the origin of
momenta, and (b) an infrared finite gluon propagator that goes to zero
at the  origin of momenta.   Although the lattice  simulations results
definitely point  out to  an infrared finite  propagator, there  is no
agreement  about its  behavior  at $q^2=0$.   A  quite recent  lattice
argument favors a non zero value for $D(0)$ \cite{cucchieri}.

The  SDE solutions for the gluon  propagator were ob\-tai\-ned
in different gauges  as well as in a  gauge independent approach. Here
we just assume  their validity for any gauge  and write the propagator
as
\begin{equation}
D_{\mu \nu} (q^2) = \left( \delta_{\mu \nu} -\eta \frac{q_{\mu} q_{\nu}}{q^2}
   \right) D (q^2)\,,
\end{equation}

\noindent  with $\eta=0$  in the  Feynman  gauge and  $\eta=1$ in  the
Landau  Gauge.  In  the  following  we briefly  describe  some of  the
solutions found in the literature.

\subsection{Infrared finite propagators with $D(0)\neq 0$}

\subsubsection{Cornwall solution $^{40}$}

This solution was  obtained by Cornwall many years  ago and predicts a
running  coupling constant  and  gluon propagator  which are  infrared
finite and non-null at the origin. It also predicts the existence of a
dynamical gluon mass which is responsible for the infrared behavior of
the  aforementioned  quantities.   This  dynamical gluon  mass  has  a
dependence on the momentum given by

\begin{equation}
 M_g^2 (q^2) = m_g^2  \left[ \frac{\mathrm{ln} \left( \frac{q^2 + 4
   m_g^2}{\Lambda^2} \right)}{\mathrm{ln} \left( \frac{4 m_g^2}{\Lambda^2}
   \right)} \right]^{- \frac{12}{11}}, 
\end{equation}

\noindent    where   $m_g$    is    the   gluon    mass   scale    and
$\Lambda=\Lambda_{\mathrm{QCD}}$  is  the  QCD  scale.   The  running  coupling
constant is

\begin{equation}
\alpha_{\mathrm{s}}^{\mathrm{Ia}} (q^2) = \frac{4 \pi}{\beta_0\mathrm{ln} \left( \frac{q^2 + 4
   M_g^2 (q^2)}{\Lambda^2} \right)},
\end{equation}

\noindent with $\beta_0 =  11-\frac{2}{3}n_f$, and $n_f$ is the number
of active quark flavors at a given scale. Although the SDE were solved
for the pure  gauge theory it is usually  assumed that their solutions
will not be strongly affected  by the introduction of fermions as long
as $n_f$ is not too large \cite{natale3,natale4}. The gluon propagator
is equal to

\begin{equation}
D_{\mathrm{Ia}}(q^2) = \frac{1}{\left[ q^2 + M_g^2 (q^2) \right] \beta_0 g^2 \ln
   \left( \frac{q^2 +4 M_g^2 (q^2)}{\Lambda^2} \right)}\,,
\end{equation}
\noindent where  the parameter $g^2$  is the strong coupling  given by
$\alpha_s=g^2/4\pi$. The  dynamical gluon  mass $m_g  $  has typical
values \cite{natale3}-\cite{natale5,cornwall}:
\begin{equation} m_g = 500 \pm 200\, \mathrm{MeV} \,\,\, .
\end{equation}

\subsubsection{Fit of Ref.~50}

Another   solution  with  a   dynamical  gluon   mass  was   found  in
Ref.~\cite{aguilar1}. This  solution was fitted by the  simplest way to
have a  dynamically massive gluon,  whose mass obeys the  standard OPE
behavior at high energy ($m_g^2 (q^2) \propto m_0^4/q^2$)
\begin{equation}
 D_{\mathrm{IIa}} (q^2) = \frac{1}{q^2 + M^2(q^2)}, 
\end{equation}
\noindent with  $m_0$ being the  gluon mass scale, and  $M^2(q^2)$ the
dynamical gluon mass given by
\begin{equation}
M^2(q^2) = \frac{m_0^4}{q^2+m_0^2}. 
\end{equation}
The  coupling  constant  is  similar  to the  previous  solution,  with
$M_g(q^2)$ being replaced by $M^2(q^2)$.

\subsubsection{Aguilar-Papavassiliou solution $^{51,52}$}

The SDE solution obtained by  Aguilar and Papavassiliou results from a
quite  detailed  analysis of  the  gluon  propagator  using the  pinch
technique, which maintains the desirable property of transversality of
the propagator. This solution is represented by
\begin{equation}
D_{\mathrm{IIc}} (q_{}^2) = \frac{1}{q^2 + m^2 (q^2)}\,. 
\end{equation}

\noindent The dynamical mass is:
\begin{equation}
m^2 (q^{^2}) = \frac{m_0^4}{q^2 +  m_0^2} \left[ \ln \left( \frac{q^2 + \rho
   m^2_0}{\Lambda^2} \right)/\ln\left(\frac{\rho m_0^2}{\Lambda^2}\right) \right]^{\gamma_2 - 1}\,,
\end{equation}

\noindent with $\gamma_2 > 1$. The running coupling constant is given by
\begin{equation}
g^2(q^2) = \left[ \beta_0 \ln \left( \frac{q^2 + f (q^2, m^2 (q^2))}{\Lambda^2}
   \right) \right]^{- 1}\,,
\end{equation}
\noindent where the function $f (q^2,  m^2 (q^2))$ is given by a power
law expression
\begin{eqnarray}
 f (q^2, m^2 (q^2))& = & \rho_1 m^2 (q^2) + \rho_2 \frac{m^4 (q^2)}{q^2 + m^2
   (q^2)} 
 +  \rho^{}_3 \frac{m^6 (q^2)}{[q^2 + m^2 (q^2)]^2}\,.
\end{eqnarray}

\noindent  The  parameters $\rho$,  $\rho_1$,  $\rho_2$, $\rho_3$  and
$\gamma_2$ are fitted  numerically.  For $\Lambda = 300$  MeV and $m_0
=250\,\mathrm{MeV}$  the numerical  values of  the parameters  are the
following: $\rho = 2.47$, $\gamma_2  = 1.76$ $\rho_1 = 5.615$, $\rho_2
= -8.523$,  $\rho_3 =  3.584$; and for  $m_0 =  600\,\mathrm{MeV}$ the
parameters    are:   $\rho=1.234$,    $\gamma=1.64$,   $\rho_1=2.894$,
$\rho_2=-4.534$, $\rho_3=2.043$.

\begin{figure}[pt]
\includegraphics[width=10cm]{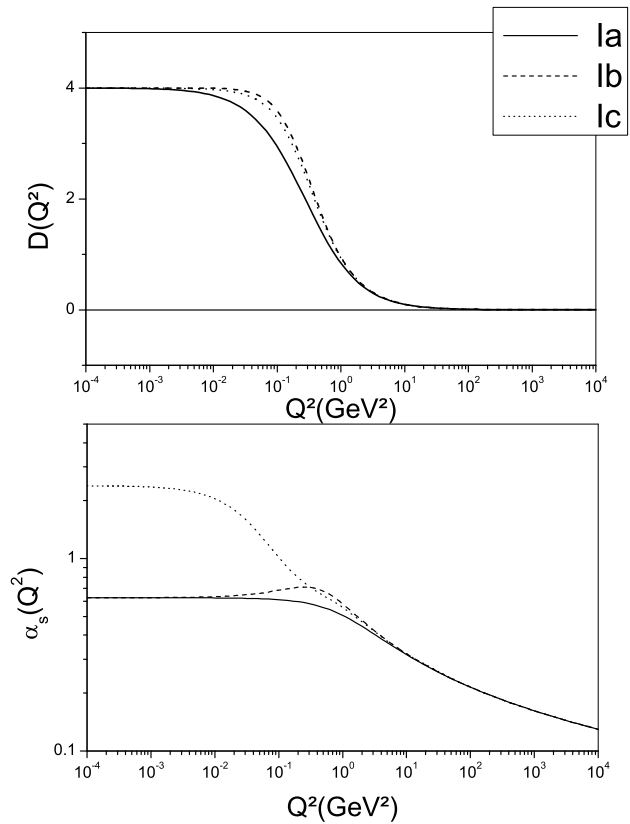}
\vspace*{8pt}
\caption{\label{prop1}  Coupling  constants  and propagators  for  the
  three solutions  with infrared finite propagators  and $D(0)\neq 0$,
  with $m_g = m_0 = 500$ MeV: (Ia) Cornwall solution, (Ib) solution of
  Ref.~50, (Ic) Aguilar-Papavassiliou solution.}
\end{figure}

The Fig.~\ref{prop1}  shown the behavior  of the propagators and
coupling constants of the three solutions presented in this subsection.

\subsection{Infrared finite propagators with $D(0)= 0$}

\subsubsection{Alkofer {\it et al.} solution $^{53}$}

This solution, obtained in the Landau gauge, has a different qualitative
behavior relative to the ones we have presented so far. Although
this solution also predicts a freezing of the coupling constant, and
the solution for the gluon propagator is also infrared finite, the
propagator vanishes at the origin of momenta.  The behavior of the
running coupling constant is given by
\begin{equation}
 \alpha_{\mathrm{s}}^{\mathrm{IIa}} (q^2) = \frac{\alpha_A (0)}{\ln [e + a_1 (q^2)^{a_2} +
   b_1 (q^2)^{b 2}]}\,,
\end{equation}

\noindent and the propagator is equal to
\begin{equation}\label{propalk} 
D_{\mathrm{IIa}} (q^2) = \frac{Z (q^2)}{q^2}\,, 
\end{equation}
\noindent  where  $Z  (q^2)$ and $R(q^2)$ are  fitted  as
\begin{equation}\label{Z}
 Z (q^2) = \left( \frac{\alpha_{\mathrm{sA}} (q^2)}{\alpha_{\mathrm{sA}} (\mu)}
   \right)^{1 + 2 \delta} R^2 (q^2)\,,
\end{equation}
and
\begin{equation}
  R (q^2) = \frac{c (q^2)^{\kappa} + d (q^2)^{2 \kappa}}{1 + c (q^2)^{\kappa}
   + d (q^2)^{2 \kappa}}\,. 
\end{equation}
\noindent The constants are parameters  obtained in the fitting of the
SDE numerical solution: 

 $\alpha_A (0) = 2.972$,  $\alpha_{\mathrm{sA}} (\mu) = 0.9676$,  
 $a_1 = 5.292\, \mathrm{GeV}^{- 2  a_2}$, 

 $ b_1 =0.034\, \mathrm{GeV}^{- 2
    b_2}$,
$a_2 = 2.324$,  $b_2 = 3.169$,

$\kappa = 0.5953$,  $\delta = - 9 / 4$,
$c = 1.8934$ GeV$^{- 2 \kappa}$,  

$d = 4.6944$ GeV$^{- 4 \kappa}.$

\subsubsection{Fit of Ref.~54}

A more recent  fit obtained by Alkofer {\it  et al.} \cite{alkofer2,alkoferc},
resulting  from different  approximations for  the SDE,  leads  to the
following expression  for the function  $Z (q^2)$ that appears  in the
equation (\ref{propalk})
\begin{equation} Z_{\mathrm{fit}} (q^2) = w\left(  \frac{q^2}{q^2 + \Lambda^2} \right)^{2
   \kappa} (\alpha_{\mathrm{fit}} (q^2))^{- \gamma}\,,
\end{equation}
\noindent where $\gamma = (- 13 N_C +  4 n_f) / (22 N_C - 4 n_f)$. The
the  parameter  $w$  comes  from  the normalization  of  the  function
$Z(p^2)$ as  $Z(p^2=1 GeV)=1$,  giving $w=1.32$ for  $\Lambda=300$ MeV
and $w=1.4$  for $\Lambda=225$ MeV.  The propagator  for this solution
is then
\begin{equation}\label{propalk2} 
D_{\mathrm{IIb}} (q^2) = \frac{Z_{\mathrm{fit}} (q^2)}{q^2} \,,
\end{equation}
and the running coupling constant is given
by
\begin{eqnarray} 
\alpha_{\mathrm{fit}} (q^2)  =   \frac{\alpha_{S (0)}}{1 + q^2 / \Lambda^2} +
   \frac{4   \pi}{\beta_0}    \frac{q^2}{q^2   +   \Lambda^2} \times  \bigg(
   \frac{1}{\ln  (q^2 / \Lambda^2)}
  -  \frac{1}{q^2  / \Lambda^2  - 1}\bigg)\,,\nonumber
\end{eqnarray}
\noindent with $\beta_0 = (11 N_C - 2 n_f) / 3$.

\begin{figure}[pt]
\includegraphics[width=10cm]{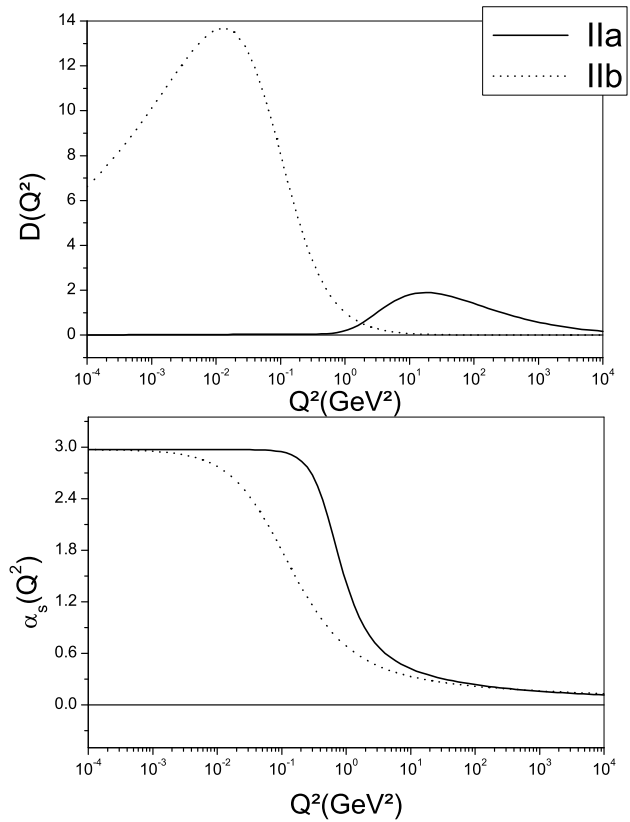}
\vspace*{8pt}
\caption{\label{prop2} Coupling constants and propagators for the two
  solutions with infrared null propagators: (IIa) Alkofer {\it et al.}
  solution, (IIb) solution of Ref.~51.}
\end{figure}

The Fig.~\ref{prop2}  shows the behavior  of the propagators and
coupling constants of the two solutions presented in this subsection.

\section{Numerical results}

In the numerical calculations we use the asymptotic expression for the
wave functions of the light me\-sons, pions and kaons,
\begin{equation}
\Phi_{\pi,K}(x)\,=\,6\,x\,(1-x),
\end{equation}
while for the $D$ mesons we use
\begin{equation}
\Phi_D(x)\,=\,6\,x\,(1-x)\,[1+a_D(1-2x)],
\end{equation}
with  $a_D=0.8$ for  $D^\pm$  and $a_D=0.3$  for  the $D_s^\pm$  meson
\cite{lu1,lu2}. We  neglected the momentum of  the ``spectator'' quark
in  the  calculation,   i.e.  we  assume  $x,y\gg  z$,   and  in  this
approximation the distribution amplitude for the $B$ meson is given by
a  delta function. Some  results will  also be  presented in  the case
where  the  wave functions  of  light  mesons  are represented  by  an
expansion  in  Gegenbauer polynomials.   The  Wilson coefficients  are
computed   using   the   equations   given  in   the   appendices   of
Ref.~\cite{wilson}.    We  do  not  include  strong   phases  in  the
calculation of the integrals, i.e.   our amplitudes will be real apart
from CKM phases.  It is known that the strong  phases generated by the
annihilation  amplitudes are  crucial for  predicting  CP asymmetries.
For   example,  in   the  Ref.~\cite{yangb}   the  strong   phases  of
annihilation amplitudes  are generated by applying  the Cutkosky rules
for the quark propagators.  However, the asymmetries are obtained from
the  ratio of  amplitudes  and  the CP  asymmetry  parameters are  not
sensitive  to  the  cutoff   effect  (or  the  infrared  finite  gluon
propagator).  Moreover the main  theoretical errors originate from the
CKM parameters \cite{yangb}.

We also use the following parameters \cite{pdg}:
\begin{itemize}
 \item Masses:  
\begin{quote}$m_{B_d}=5.28$  GeV, $m_{B_s}=5.37$ GeV,
  $m_D=1.87		$ GeV,  $m_{D_s}=1.97$ GeV,
 $m_b=4.7$ GeV;\end{quote}
\item Decay constants:
\begin{quote} $f_{B_d}=0.200$  GeV, $f_{B_s}=0.236$ GeV,
  $f_{D}=0.226$ GeV,  $f_{D_s}=0.241$  GeV,
  $f_{\pi}=0.132$  GeV,  $f_K=0.160$ GeV;
\end{quote}
\item Lifetimes: 
\begin{quote}
$\tau_{B_d}=1.54$ ps, $\tau_{B_s}=1.466$ ps; \end{quote}
\item CKM parameters:
\begin{quote}
$A=0.818$, $\lambda=0.2272$, $\bar{\rho}=0.221$, 
$\bar{\eta}=0.340$.\end{quote}
\end{itemize}

The Table  \ref{table1} shows the  branching ratios obtained  for each
decay  channel for  the different  propagators and  coupling constants
discussed  in  the  previous  section.   As we  shall  detail  in  the
conclusions, it  is possible to  see that the non-leptonic  $B$ decays
resulting from annihilation channels are sensitive to the infrared QCD
behavior.

\begin{table}[t!]
  \caption{Branching  ratios $\mathcal{B}$ for  decays of
    the  $B$ meson  obtained with  the different  infrared  finite gluon
    propagators and coupling  constants, and with $m_g =  m_0 = 500$ MeV
    and   $\mu  =   m_b$:  Cornwall   (Ia),   Ref.~50  (Ib),
    Aguilar-Papavassiliou   (Ic),   Alkofer   {\it   et   al.}    (IIa),
    Ref.~54 (IIb).  We use $\Lambda_{\mathrm{QCD}} = 225$
    Mev, for $\mu\leq m_b$ and $\Lambda_{\mathrm{QCD}} = 300$ Mev, for
    $\mu>m_b$. The experimental data for the branching ratios of the decay channels are also shown,  and the corresponding references. \label{table1}}
  {\begin{tabular}{llllllcc}\toprule
     ${\mathcal{B}}$(Decay channel) & $D_{Ia}$            & $D_{Ib}$    &  $D_{Ic}$ & $D_{IIa}$ & $D_{IIb}$ & Experimental data & Ref. \\\colrule
      \footnotesize{${\mathcal B}(B_s^0${{$\rightarrow$}} $\pi^+\pi^-)${{$\times$}}$10^7$} & 1.08 & 1.58 & 1.02 & 4.30  & 3.73 &  $<17$ & \cite{cdf1}\\
      &   &  & &  & &  $<13.6$ & \cite{cdf2}\\
${\mathcal B}(B_d^{0}${{$\rightarrow$}}$K^+K^-)${{$\times$}}$10^8$ & 4.92 & 7.18 & 4.63 & 19.62 & 16.90   & $9 ^{+18}_{-13} {\pm} 1$ & \cite{belle}\\
& & & & & & $<37$ & \cite{pdg}\\
& & & & & & $4{\pm}15{\pm}8$ & \cite{babar}\\

${\mathcal B}(B_s^0${$\rightarrow$}$D^-\pi^+)${{$\times$}}$10^6$ & 1.03 &  1.54 & 0.99 & 4.40  &  4.02 & -- & -- \\
${\mathcal B}(B_s^0${{$\rightarrow$}}$D^+\pi^-)${{$\times$}}$10^7$ & 1.28 & 1.88 & 1.22 & 5.31 & 4.76 & -- & --
 \\
${\mathcal B}(B_d^0${{$\rightarrow$}}$D_s^-K^+)${{$\times$}}$10^5$ & 1.34 & 1.98 & 1.27 & 5.61 & 3.72   & $4.6 ^{+1.2}_{-1.1} {\pm} 1.0 $ &  \cite{belle2}\\ 
& & & & & & $3.2\pm 1.0\pm 1.0$&  \cite{babar2} \\

& & & & &  & $3.1\pm 0.8$ &\cite{pdg} \\
${\mathcal B}(B^0_d${{$\rightarrow$}}$ D_s^+K^-)${{$\times$}}$10^8$ & 0.67 & 0.99 & 0.64 & 2.66 & 1.79 & $<1.1 \times 10^{5}$ & \cite{pdg}  \\ \botrule
\end{tabular}}
\end{table}

\section{Conclusions}

We  have  studied  some   decay  channels  of  $B^0_s\,(\bar{B}^0)$  and
$B^0_d\,(\bar{B}_d^0)$  mesons  which  occur  through  the  annihilation
diagrams:   $B_s^0\to  \pi^+\pi^-,\,D^{\pm}\pi^{\mp}$   and  $B_d^0\to
K^+K^-,\,D_s^{\pm}K^{\mp}$. We have  argued that infrared finite gluon
propagators and  running coupling  constants obtained as  solutions of
the QCD Schwinger-Dyson  equations, may serve as a  natural cutoff for
the  end-point divergences  that appear  in the  calculation  of these
decays.

We  computed  several  branching  ratios  for some  of  the  different
solutions  of gluon propagators  and coupling  constants found  in the
literature.   These  different   solutions  appear  due  to  different
approximations performed to sol\-ve the  SDE. Our results are shown in
Table \ref{table1}. We argue that a systematic study of $B$ decays may
help to give  information about the infrared QCD  behavior.  Note that
the results of  Table \ref{table1} are not trivial,  in the sense that
the differences in the branching  ratios come out from the integration
over  different expressions  for the  gluon propagators  multiplied by
different wave functions.  It is possible to see in Table \ref{table1}
that  the different  classes of  gluon propagators,  according  to its
infrared behavior, lead  to branching ratios differing by  a factor of
approximately  2 to  4. The  existing experimental  data of  the decay
channels that we  have discussed is shown in  Table \ref{table1}.  The
experimental  data,   when  confronted  with  the   results  of  Table
\ref{table1}, may already be used to claim that some approximations to
solve the SDE  of pure gauge QCD lead to poor  predictions for some of
the branching  ratios (see, for  instance, the results related  to the
propagator indicated by $D_{IIa}$).

In  the  calculation of  B  mesons  decays  through the  factorization
theorems, the  dominant sources  of theoretical uncertainties  are the
mesons  wave functions  and the  scale  parameter $\mu$,  as has  been
pointed  out  in   Ref.~\cite{beneke2}.   We  have  analyzed  these
uncertainties  in our  calculations  for the  decay channel  $B^0_s\to
D^-_sK^+$ and our  results are shown in Table  \ref{table2}, and as we
shall see in  the sequence all these uncertainties  are less important
than  the   differences  originated  by  the  two   classes  of  gluon
propagators.   We can  see  that  the branching  ratios  are not  very
sensitive to  the shape  of the wave  functions, with a  difference of
about 2\%  between the results  using the asymptotic function  and the
expansion in  Gegenbauer polynomials, which is much  below the present
experimental accuracy.   The main source  of uncertainty is  the scale
parameter, with a variation  of about 40-60\%.  This strong dependence
is   expected  since   the   annihilation  decay   processes  are   of
${\mathcal{O}}(\alpha_s(\mu))$  in the  perturbative  expansion.  This
dependence  can  be reduced  considering  higher  order  terms of  the
perturbative  expansion.  Once the  uncertainties in  these quantities
become well known we certainly  will be able to extract information on
the  gluon  mass  parameter.   It  is important  to  notice  that  the
amplitude that we  calculate for each specific decay  is a convolution
of the gluon propagator and  coupling constant with the wave functions
of the final state mesons, therefore the result is peaked at different
momenta, depending on  the meson masses, and help  to discriminate the
form  of   the  gluon  propagator  and   infrared  coupling  constant.
Actually, this is the reason for the different sensitivities on the IR
cutoff in the  decay channels discussed in this  work, which will help
to eliminate the possible  uncertainties. It should also be remembered
that the  next order twist  can modify the  results, but if we  have a
large  confidence level  in the  values  of these  many parameters  or
functions, we  certainly can test  the infrared QCD behavior  in these
decays.  In Table  \ref{table2}  it is  also  evident the  differences
between using SDE  solutions of the type I and  II for the propagators
and coupling constants.

\begin{table}
  \caption{Branching ratios for the decay channel $B^0_s\to D^-_sK^+$,    ${\mathcal  B}(B_d^0 \rightarrow  D_s^-K^+)\times  10^5$. We compare
    two different scales, $\mu=m_b$  and $\mu=m_b/2$, and use the asymptotic wave function of the $K$ meson and also the wave function expanded in Gegenbauer polynomials (at the    scale   $\mu=m_b$,  with  $\alpha_{1}(1\,{\mathrm{GeV}})=0.17$  and  $\alpha_2(1\,{\mathrm{GeV}})=0.2$, Ref.~56).   We  also show  the
    results for different gluon masses. \label{table2} }
  {\begin{tabular}{cccc}\toprule
      & Asymptotic &             & Gegenbauer  \\
      Gluon propagators       & $\mu=m_b$ & $\mu=m_b/2$ & $\mu=m_b$  \\\colrule
      $D_{Ia}$, $m_g=400$ MeV	& 1.95	 & 3.25   & 2.00 \\
      $D_{Ia}$, $m_g=500$ MeV	& 1.34	 & 2.09    & 1.36  \\
      $D_{Ib}$, $m_0=400$ MeV   & 2.70  & 4.89   & 2.75  \\
      $D_{Ib}$, $m_0=500$ MeV   & 1.98	 & 3.48   & 2.02  \\
      $D_{Ic}$, $m_0=500$ MeV   & 1.84  & 3.10   & 1.87  \\
      $D_{Ic}$, $m_0=600$ MeV   & 1.40  &  2.40   & 1.41 \\
      $D_{IIa}$	                & 5.61	&  16.78  & 5.75 \\
      $D_{IIb}$	        & 3.72	&  7.32 &  3.95  \\\botrule
\end{tabular}}
\end{table}


It  is clearly  necessary  to collect  much  more data  to settle  the
question about  the infrared QCD  behavior, at least in  what concerns
the   infrared  behavior   of  the   gluon  propagator   and  coupling
constant. We know that the  comparison with the experimental data will
also depend  on quantities like  the distribution functions,  but with
the LHCb experiment  the uncertainty in the experimental  data will be
narrowed, and it will be  possible to explore different models until a
reasonable physical picture of the $B$ meson decays is obtained.

From the tables shown above we see that the results depend sensitively
on the  different solutions of  infrared finite gluon  propagators. It
must be said  that a lot of  progress has been done in  the subject of
SDE for  the gluon  propagators  recently, this
also happened to attract the  attention of lattice researchers, and we
believe that the  lattice simulation of the gluon  propagator can only
be fitted  nicely by the  ``massive gluon solution", see  for instance
Ref.~\cite{aguilar08}, although this is  not an unanimous opinion.  If
with  improved lattice  results we  come  to the  conclusion that  the
massive gluon propagator  is the one chosen by  Nature, we verify that
the theoretical uncertainty will be  only due to the poor knowledge of
the gluon  mass value. In this case  we can see from  our results that
the variation  of the branching  ratios is not  so large, and  will be
smaller as  long as this gluon  mass is better  determined through all
its phenomenological consequences.

In  order to  fully understand  the  non-leptonic $B$  decays that  we
discussed in this  work we see two specific  directions of study.  One
of these  is the need of  finding a gauge invariant  truncation in the
context of the  SDE. The importance of constructing  such a truncation
scheme  is the  possibility  of defining  a non-perturbative  effective
charge for  QCD, which constitutes  a generalization in  a non-abelian
theory of the QED effective charge as discussed in Ref.~\cite{bp}.  If this
step was accomplished  in all its glory, it  would be possible discard
one of  the solutions of  the infrared sector  of QCD on the  basis of
gauge-invariance.  The other direction  of study is that there should
be a systematic use of  infrared finite gluon propagators and coupling
constant  in  non-leptonic  $B$  meson  decays. Any  $B$  meson  decay
involving a gluon  exchange will be affected by the  IR cutoff that we
discussed here,  and we  intend to  study other decays  as well  as to
implement the  higher twist effects in this  calculation. The infrared
finite  gluon   propagator  provides   a  natural  cutoff   for  these
phenomenological processes.   As we  have seen in  Table \ref{table1},
the  branching ratios  of  heavy  mesons depend  on  the infrared  QCD
details, and this  will certainly happen when we  have light mesons in
the final states.  The most remarkable result of Table \ref{table1} is
that with the concept of a dynamically massive gluon it is possible to
predict the  branching ratios of  some $B$ decays compatible  with the
experimental data,  without the  help of any  {\it ad~hoc}  cutoff and
with the same mass scale that fits the experimental data of many other
processes.

\section*{Acknowledgments}

We would like to thank A.  C. Aguilar for discussions, and Ya-Dong Yang
and  Diego  Tonelli for  useful  correspondence.   This research  was
supported by the  Conselho Nacional de Desenvolvimento Cient\'{\i}fico
e Tecnol\'ogico-CNPq (AAN and CMZ).


\begin{thebibliography}{}
\expandafter\ifx\csname natexlab\endcsname\relax\def\natexlab#1{#1}\fi
\expandafter\ifx\csname bibnamefont\endcsname\relax
  \def\bibnamefont#1{#1}\fi
\expandafter\ifx\csname bibfnamefont\endcsname\relax
  \def\bibfnamefont#1{#1}\fi
\expandafter\ifx\csname citenamefont\endcsname\relax
  \def\citenamefont#1{#1}\fi
\expandafter\ifx\csname url\endcsname\relax
  \def\url#1{\texttt{#1}}\fi
\expandafter\ifx\csname urlprefix\endcsname\relax\def\urlprefix{URL }\fi
\providecommand{\bibinfo}[2]{#2}
\providecommand{\eprint}[2][]{\url{#2}}

\bibitem{fa}
\bibinfo{author}{\bibfnamefont{M.}~\bibnamefont{Wirbel}},
  \bibinfo{author}{\bibfnamefont{B.}~\bibnamefont{Stech}}, \bibnamefont{and}
  \bibinfo{author}{\bibfnamefont{B.}~\bibnamefont{Manfred}},
  \bibinfo{journal}{Z. Phys.} \textbf{\bibinfo{volume}{C29}},
  \bibinfo{pages}{637} (\bibinfo{year}{1985}).

\bibitem{fa1}
\bibinfo{author}{\bibfnamefont{M.}~\bibnamefont{Bauer}},
  \bibinfo{author}{\bibfnamefont{B.}~\bibnamefont{Stech}}, \bibnamefont{and}
  \bibinfo{author}{\bibfnamefont{M.}~\bibnamefont{Wirbels}},
  \bibinfo{journal}{Z. Phys.} \textbf{\bibinfo{volume}{C34}},
  \bibinfo{pages}{103} (\bibinfo{year}{1987}).

\bibitem{fa2}
\bibinfo{author}{\bibfnamefont{L.~L.} \bibnamefont{Chau}},
  \bibinfo{author}{\bibfnamefont{H.~Y.} \bibnamefont{Cheng}},
  \bibinfo{author}{\bibfnamefont{W.~K.} \bibnamefont{Sze}},
  \bibinfo{author}{\bibfnamefont{H.}~\bibnamefont{Yao}}, \bibnamefont{and}
  \bibinfo{author}{\bibfnamefont{B.}~\bibnamefont{Tseng}},
  \bibinfo{journal}{Phys. Rev.} \textbf{\bibinfo{volume}{D43}},
  \bibinfo{pages}{2176} (\bibinfo{year}{1991}).

\bibitem{brodsky}
\bibinfo{author}{\bibfnamefont{A.}~\bibnamefont{Szczepaniak}},
  \bibinfo{author}{\bibfnamefont{E.~M.} \bibnamefont{Henley}},
  \bibnamefont{and} \bibinfo{author}{\bibfnamefont{S.~J.}
  \bibnamefont{Brodsky}}, \bibinfo{journal}{Phys. Lett.}
  \textbf{\bibinfo{volume}{B243}}, \bibinfo{pages}{287} (\bibinfo{year}{1990}).

\bibitem{lepage}
\bibinfo{author}{\bibfnamefont{G.~P.} \bibnamefont{Lepage}} \bibnamefont{and}
  \bibinfo{author}{\bibfnamefont{S.~J.} \bibnamefont{Brodsky}},
  \bibinfo{journal}{Phys. Lett.} \textbf{\bibinfo{volume}{B87}},
  \bibinfo{pages}{359} (\bibinfo{year}{1979}).

\bibitem{lepagea}
\bibinfo{author}{\bibfnamefont{G.~P.} \bibnamefont{Lepage}} \bibnamefont{and}
  \bibinfo{author}{\bibfnamefont{S.~J.} \bibnamefont{Brodsky}},
  \bibinfo{journal}{Phys. Rev.} \textbf{\bibinfo{volume}{D22}},
  \bibinfo{pages}{2157} (\bibinfo{year}{1980}).

\bibitem{sterman}
\bibinfo{author}{\bibfnamefont{J.}~\bibnamefont{Botts}} \bibnamefont{and}
  \bibinfo{author}{\bibfnamefont{G.}~\bibnamefont{Sterman}},
  \bibinfo{journal}{Nucl. Phys.} \textbf{\bibinfo{volume}{B325}},
  \bibinfo{pages}{62} (\bibinfo{year}{1989}).

\bibitem{stermana}
\bibinfo{author}{\bibfnamefont{H.-N.} \bibnamefont{Li}} \bibnamefont{and}
  \bibinfo{author}{\bibfnamefont{G.}~\bibnamefont{Sterman}},
  \bibinfo{journal}{Nucl. Phys.} \textbf{\bibinfo{volume}{B381}},
  \bibinfo{pages}{129} (\bibinfo{year}{1992}).



\bibitem{beneke}
\bibinfo{author}{\bibfnamefont{M.}~\bibnamefont{Beneke}},
  \bibinfo{author}{\bibfnamefont{G.}~\bibnamefont{Buchalla}},
  \bibinfo{author}{\bibfnamefont{M.}~\bibnamefont{Neubert}}, \bibnamefont{and}
  \bibinfo{author}{\bibfnamefont{C.~T.} \bibnamefont{Sachrajda}},
  \bibinfo{journal}{Phys. Rev. Lett.} \textbf{\bibinfo{volume}{83}},
  \bibinfo{pages}{1914} (\bibinfo{year}{1999}), \eprint{hep-ph/9905312}.

\bibitem{beneke1}
\bibinfo{author}{\bibfnamefont{M.}~\bibnamefont{Beneke}},
  \bibinfo{author}{\bibfnamefont{G.}~\bibnamefont{Buchalla}},
  \bibinfo{author}{\bibfnamefont{M.}~\bibnamefont{Neubert}}, \bibnamefont{and}
  \bibinfo{author}{\bibfnamefont{C.~T.} \bibnamefont{Sachrajda}},
  \bibinfo{journal}{Nucl. Phys.} \textbf{\bibinfo{volume}{B591}},
  \bibinfo{pages}{313} (\bibinfo{year}{2000}), \eprint{hep-ph/0006124}.

\bibitem{beneke2}
\bibinfo{author}{\bibfnamefont{M.}~\bibnamefont{Beneke}},
  \bibinfo{author}{\bibfnamefont{G.}~\bibnamefont{Buchalla}},
  \bibinfo{author}{\bibfnamefont{M.}~\bibnamefont{Neubert}}, \bibnamefont{and}
  \bibinfo{author}{\bibfnamefont{C.~T.} \bibnamefont{Sachrajda}},
  \bibinfo{journal}{Nucl. Phys.} \textbf{\bibinfo{volume}{B606}},
  \bibinfo{pages}{245} (\bibinfo{year}{2001}), \eprint{hep-ph/0104110}.

\bibitem{beneke3}
\bibinfo{author}{\bibfnamefont{M.}~\bibnamefont{Beneke}} \bibnamefont{and}
  \bibinfo{author}{\bibfnamefont{M.}~\bibnamefont{Neubert}},
  \bibinfo{journal}{Nucl. Phys.} \textbf{\bibinfo{volume}{B675}},
  \bibinfo{pages}{333} (\bibinfo{year}{2003}), \eprint{hep-ph/0308039}.

\bibitem{buras}
\bibinfo{author}{\bibfnamefont{G.}~\bibnamefont{Buchalla}},
  \bibinfo{author}{\bibfnamefont{A.~J.} \bibnamefont{Buras}}, \bibnamefont{and}
  \bibinfo{author}{\bibfnamefont{M.~E.} \bibnamefont{Lautenbacher}},
  \bibinfo{journal}{Rev. Mod. Phys.} \textbf{\bibinfo{volume}{68}},
  \bibinfo{pages}{1125} (\bibinfo{year}{1996}), \eprint{hep-ph/9512380}.

\bibitem{natale}
\bibinfo{author}{\bibfnamefont{A.~A.}~\bibnamefont{Natale}},
  \bibinfo{journal}{Braz. J. Phys.} \textbf{\bibinfo{volume}{37}},
  \bibinfo{pages}{306} (\bibinfo{year}{2007}), \eprint{hep-ph/ 0610256}.

\bibitem{bp}
\bibinfo{author}{\bibfnamefont{D.}~\bibnamefont{Binosi}} \bibnamefont{and}
  \bibinfo{author}{\bibfnamefont{J.}~\bibnamefont{Papavassiliou}}
  (\bibinfo{year}{2007}), \eprint{arXiv:0712.2707 [hep-ph]}.

\bibitem{lattice}
\bibinfo{author}{\bibfnamefont{A.}~\bibnamefont{Cucchieri}} \bibnamefont{and}
  \bibinfo{author}{\bibfnamefont{T.}~\bibnamefont{Mendes}}
  (\bibinfo{year}{2007}{\natexlab{a}}), \eprint{arXiv:0710.0412 [hep-lat]}.

\bibitem{bogolubsky}
\bibinfo{author}{\bibfnamefont{I.~L.} \bibnamefont{Bogolubsky}},
  \bibinfo{author}{\bibfnamefont{E.~M.} \bibnamefont{Ilgenfritz}},
  \bibinfo{author}{\bibfnamefont{M.}~\bibnamefont{Muller-Preussker}},
  \bibnamefont{and} \bibinfo{author}{\bibfnamefont{A.}~\bibnamefont{Sternbeck}}
  (\bibinfo{year}{2007}), \eprint{arXiv:0710.1968 [hep-lat]}.

\bibitem{cucchieri}
\bibinfo{author}{\bibfnamefont{A.}~\bibnamefont{Cucchieri}} \bibnamefont{and}
  \bibinfo{author}{\bibfnamefont{T.}~\bibnamefont{Mendes}}
  (\bibinfo{year}{2007}{\natexlab{b}}), \eprint{arXiv:0712.3517 [hep-lat]}.

\bibitem{natale3}
\bibinfo{author}{\bibfnamefont{F.}~\bibnamefont{Halzen}},
  \bibinfo{author}{\bibfnamefont{G.~I.} \bibnamefont{Krein}}, \bibnamefont{and}
  \bibinfo{author}{\bibfnamefont{A.~A.} \bibnamefont{Natale}},
  \bibinfo{journal}{Phys. Rev.} \textbf{\bibinfo{volume}{D47}},
  \bibinfo{pages}{295} (\bibinfo{year}{1993}).

\bibitem{natale3a}
\bibinfo{author}{\bibfnamefont{M.~B.~G.} \bibnamefont{Ducati}},
  \bibinfo{author}{\bibfnamefont{F.}~\bibnamefont{Halzen}}, \bibnamefont{and}
  \bibinfo{author}{\bibfnamefont{A.~A.} \bibnamefont{Natale}},
  \bibinfo{journal}{Phys. Rev.} \textbf{\bibinfo{volume}{D48}},
  \bibinfo{pages}{2324} (\bibinfo{year}{1993}), \eprint{hep-ph/9304276}.

\bibitem{natale4}
\bibinfo{author}{\bibfnamefont{A.~C.} \bibnamefont{Aguilar}},
  \bibinfo{author}{\bibfnamefont{A.}~\bibnamefont{Mihara}}, \bibnamefont{and}
  \bibinfo{author}{\bibfnamefont{A.~A.} \bibnamefont{Natale}},
  \bibinfo{journal}{Phys. Rev.} \textbf{\bibinfo{volume}{D65}},
  \bibinfo{pages}{054011} (\bibinfo{year}{2002}), \eprint{hep-ph/0109223}.

\bibitem{natale5}
\bibinfo{author}{\bibfnamefont{A.}~\bibnamefont{Mihara}} \bibnamefont{and}
  \bibinfo{author}{\bibfnamefont{A.~A.} \bibnamefont{Natale}},
  \bibinfo{journal}{Phys. Lett.} \textbf{\bibinfo{volume}{B482}},
  \bibinfo{pages}{378} (\bibinfo{year}{2000}), \eprint{hep-ph/0004236}.

\bibitem{natale5a}
\bibinfo{author}{\bibfnamefont{A.~C.} \bibnamefont{Aguilar}},
  \bibinfo{author}{\bibfnamefont{A.}~\bibnamefont{Mihara}}, \bibnamefont{and}
  \bibinfo{author}{\bibfnamefont{A.~A.} \bibnamefont{Natale}},
  \bibinfo{journal}{Int. J. Mod. Phys.} \textbf{\bibinfo{volume}{A19}},
  \bibinfo{pages}{249} (\bibinfo{year}{2004}).

\bibitem{natale5b}
\bibinfo{author}{\bibfnamefont{E.~G.~S.} \bibnamefont{Luna}},
  \bibinfo{author}{\bibfnamefont{A.~F.} \bibnamefont{Martini}},
  \bibinfo{author}{\bibfnamefont{A.~M.} \bibnamefont{M.~J.~Menon}},
  \bibnamefont{and} \bibinfo{author}{\bibfnamefont{A.~A.}
  \bibnamefont{Natale}}, \bibinfo{journal}{Phys. Rev.}
  \textbf{\bibinfo{volume}{D72}}, \bibinfo{pages}{034019}
  (\bibinfo{year}{2005}), \eprint{hep-ph/0507057}.

\bibitem{natale5c}
\bibinfo{author}{\bibfnamefont{E.~G.~S.} \bibnamefont{Luna}} \bibnamefont{and}
  \bibinfo{author}{\bibfnamefont{A.~A.} \bibnamefont{Natale}},
  \bibinfo{journal}{Phys. Rev.} \textbf{\bibinfo{volume}{D73}},
  \bibinfo{pages}{074019} (\bibinfo{year}{2006}), \eprint{hep-ph/0602181}.

\bibitem{natale5d}
\bibinfo{author}{\bibfnamefont{F.}~\bibnamefont{Carvalho}},
  \bibinfo{author}{\bibfnamefont{A.~A.} \bibnamefont{Natale}},
  \bibnamefont{and} \bibinfo{author}{\bibfnamefont{C.~M.}
  \bibnamefont{Zanetti}}, \bibinfo{journal}{Mod. Phys. Lett.}
  \textbf{\bibinfo{volume}{A21}}, \bibinfo{pages}{3021} (\bibinfo{year}{2006}),
  \eprint{hep-ph/0510172}.

\bibitem{natale5e}
\bibinfo{author}{\bibfnamefont{E.~G.~S.} \bibnamefont{Luna}},
  \bibinfo{author}{\bibfnamefont{A.~A.} \bibnamefont{Natale}},
  \bibnamefont{and} \bibinfo{author}{\bibfnamefont{C.~M.}
  \bibnamefont{Zanetti}}, \bibinfo{journal}{Int. J. Mod. Phys.}
  \textbf{\bibinfo{volume}{A23}}, \bibinfo{pages}{151 } (\bibinfo{year}{2008}),
  \eprint{hep-ph/0605338}.

\bibitem{brodsky2}
\bibinfo{author}{\bibfnamefont{S.~J.} \bibnamefont{Brodsky}},
  \bibinfo{author}{\bibfnamefont{C.~R.} \bibnamefont{Ji}},
  \bibinfo{author}{\bibfnamefont{A.}~\bibnamefont{Pang}}, \bibnamefont{and}
  \bibinfo{author}{\bibfnamefont{D.~G.} \bibnamefont{Robertson}},
  \bibinfo{journal}{Phys. Rev.} \textbf{\bibinfo{volume}{D57}},
  \bibinfo{pages}{245} (\bibinfo{year}{1998}), \eprint{hep-ph/9705221}.

\bibitem{apt}
\bibinfo{author}{\bibfnamefont{D.~V.} \bibnamefont{Shirkov}} \bibnamefont{and}
  \bibinfo{author}{\bibfnamefont{I.~L.} \bibnamefont{Solovtsov}},
  \bibinfo{journal}{Phys. Rev. Lett.} \textbf{\bibinfo{volume}{79}},
  \bibinfo{pages}{1209} (\bibinfo{year}{1997}), \eprint{hep-ph/9704333}.

\bibitem{apta}
\bibinfo{author}{\bibfnamefont{A.~C.} \bibnamefont{Aguilar}},
  \bibinfo{author}{\bibfnamefont{A.~V.} \bibnamefont{Nesterenko}},
  \bibnamefont{and}
  \bibinfo{author}{\bibfnamefont{J.}~\bibnamefont{Papavassiliou}},
  \bibinfo{journal}{J. Phys.} \textbf{\bibinfo{volume}{G31}},
  \bibinfo{pages}{997} (\bibinfo{year}{2005}), \eprint{hep-ph/0504195}.

\bibitem{aptb}
\bibinfo{author}{\bibfnamefont{A.~V.} \bibnamefont{Nesterenko}}
  \bibnamefont{and}
  \bibinfo{author}{\bibfnamefont{J.}~\bibnamefont{Papavassiliou}},
  \bibinfo{journal}{Int. J. Mod. Phys.} \textbf{\bibinfo{volume}{A20}},
  \bibinfo{pages}{4622} (\bibinfo{year}{2005}{\natexlab{a}}),
  \eprint{hep-ph/0409220}.

\bibitem{aptc}
\bibinfo{author}{\bibfnamefont{A.~V.} \bibnamefont{Nesterenko}}
  \bibnamefont{and}
  \bibinfo{author}{\bibfnamefont{J.}~\bibnamefont{Papavassiliou}},
  \bibinfo{journal}{Phys. Rev.} \textbf{\bibinfo{volume}{D71}},
  \bibinfo{pages}{016009} (\bibinfo{year}{2005}{\natexlab{b}}),
  \eprint{hep-ph/0410406}.

\bibitem{natale2}
\bibinfo{author}{\bibfnamefont{A.~C.} \bibnamefont{Aguilar}},
  \bibinfo{author}{\bibfnamefont{A.~A.} \bibnamefont{Natale}},
  \bibnamefont{and} \bibinfo{author}{\bibfnamefont{P.~S.~R.}
  \bibnamefont{da~Silva}}, \bibinfo{journal}{Phys. Rev. Lett.}
  \textbf{\bibinfo{volume}{90}}, \bibinfo{pages}{152001}
  (\bibinfo{year}{2003}), \eprint{hep-ph/0212105}.

\bibitem{pagels}
\bibinfo{author}{\bibfnamefont{H.}~\bibnamefont{Pagels}} \bibnamefont{and}
  \bibinfo{author}{\bibfnamefont{S.}~\bibnamefont{Stokar}},
  \bibinfo{journal}{Phys. Rev.} \textbf{\bibinfo{volume}{D20}},
  \bibinfo{pages}{2947} (\bibinfo{year}{1979}).

\bibitem{brodsky3}
\bibinfo{author}{\bibfnamefont{S.~J.} \bibnamefont{Brodsky}},
  \bibinfo{author}{\bibfnamefont{E.}~\bibnamefont{Gardi}},
  \bibinfo{author}{\bibfnamefont{G.}~\bibnamefont{Grunberg}}, \bibnamefont{and}
  \bibinfo{author}{\bibfnamefont{J.}~\bibnamefont{Rathsman}},
  \bibinfo{journal}{Phys. Rev.} \textbf{\bibinfo{volume}{D63}},
  \bibinfo{pages}{094017} (\bibinfo{year}{2001}), \eprint{hep-ph/0002065}.

\bibitem{brodsky3a}
\bibinfo{author}{\bibfnamefont{S.~J.} \bibnamefont{Brodsky}},
  \bibinfo{author}{\bibfnamefont{C.~M.} \bibnamefont{S.~Menke}},
  \bibnamefont{and} \bibinfo{author}{\bibfnamefont{J.}~\bibnamefont{Rathsman}},
  \bibinfo{journal}{Phys. Rev.} \textbf{\bibinfo{volume}{D67}},
  \bibinfo{pages}{055008} (\bibinfo{year}{2003}), \eprint{hep-ph/0212078}.

\bibitem{cornwall2007}
\bibinfo{author}{\bibfnamefont{J.~M.} \bibnamefont{Cornwall}},
  \bibinfo{journal}{Phys. Rev.} \textbf{\bibinfo{volume}{D76}},
  \bibinfo{pages}{025012} (\bibinfo{year}{2007}), \eprint{hep-th/ 0702054}.

\bibitem{brodsky4}
\bibinfo{author}{\bibfnamefont{S.~J.} \bibnamefont{Brodsky}},
  \bibinfo{journal}{Acta Phys. Polon.} \textbf{\bibinfo{volume}{B32}},
  \bibinfo{pages}{4013} (\bibinfo{year}{2001}), \eprint{hep-ph/0111340}.

\bibitem{brodsky4a}
\bibinfo{author}{\bibfnamefont{S.~J.} \bibnamefont{Brodsky}},
  \bibinfo{journal}{Fortsch. Phys.} \textbf{\bibinfo{volume}{50}},
  \bibinfo{pages}{503} (\bibinfo{year}{2002}), \eprint{hep-th/ 0111241}.

\bibitem{cornwall}
\bibinfo{author}{\bibfnamefont{J.~M.} \bibnamefont{Cornwall}},
  \bibinfo{journal}{Phys. Rev.} \textbf{\bibinfo{volume}{D26}},
  \bibinfo{pages}{1453} (\bibinfo{year}{1982}).

\bibitem{yang}
\bibinfo{author}{\bibfnamefont{S.}~\bibnamefont{Bar-Shalom}},
  \bibinfo{author}{\bibfnamefont{G.}~\bibnamefont{Eilam}}, \bibnamefont{and}
  \bibinfo{author}{\bibfnamefont{Y.~D.} \bibnamefont{Yang}},
  \bibinfo{journal}{Phys. Rev.} \textbf{\bibinfo{volume}{D67}},
  \bibinfo{pages}{014007} (\bibinfo{year}{2003}), \eprint{hep-ph/0201244}.

\bibitem{yanga}
\bibinfo{author}{\bibfnamefont{Y.~D.} \bibnamefont{Yang}},
  \bibinfo{author}{\bibfnamefont{F.}~\bibnamefont{Su}},
  \bibinfo{author}{\bibfnamefont{G.~R.} \bibnamefont{Lu}}, \bibnamefont{and}
  \bibinfo{author}{\bibfnamefont{H.~J.} \bibnamefont{Hao}},
  \bibinfo{journal}{Eur. Phys. J.} \textbf{\bibinfo{volume}{C44}},
  \bibinfo{pages}{243} (\bibinfo{year}{2005}), \eprint{hep-ph/0507326}.

\bibitem{yangb}
\bibinfo{author}{\bibfnamefont{Y.~L.~W.} \bibnamefont{F.~Su}},
  \bibinfo{author}{\bibfnamefont{Y.~D.} \bibnamefont{Yang}}, \bibnamefont{and}
  \bibinfo{author}{\bibfnamefont{C.}~\bibnamefont{Zhuang}},
  \bibinfo{journal}{Eur. Phys. J.} \textbf{\bibinfo{volume}{C48}},
  \bibinfo{pages}{401} (\bibinfo{year}{2006}), \eprint{hep-ph/0604082}.


\bibitem{pqcd}
\bibinfo{author}{\bibfnamefont{H.~N.} \bibnamefont{Li}} \bibnamefont{and}
  \bibinfo{author}{\bibfnamefont{H.~L.} \bibnamefont{Yu}},
  \bibinfo{journal}{Phys. Rev. Lett.} \textbf{\bibinfo{volume}{74}},
  \bibinfo{pages}{4388} (\bibinfo{year}{1995}{\natexlab{a}}),
  \eprint{hep-ph/9409313}.

\bibitem{pqcd1}
\bibinfo{author}{\bibfnamefont{H.~N.} \bibnamefont{Li}} \bibnamefont{and}
  \bibinfo{author}{\bibfnamefont{H.~L.} \bibnamefont{Yu}},
  \bibinfo{journal}{Phys. Lett.} \textbf{\bibinfo{volume}{B353}},
  \bibinfo{pages}{301} (\bibinfo{year}{1995}{\natexlab{b}}).

\bibitem{pqcd2}
\bibinfo{author}{\bibfnamefont{Y.~Y.} \bibnamefont{Keum}},
  \bibinfo{author}{\bibfnamefont{H.~N.} \bibnamefont{Li}}, \bibnamefont{and}
  \bibinfo{author}{\bibfnamefont{A.~I.} \bibnamefont{Sanda}},
  \bibinfo{journal}{Phys. Rev.} \textbf{\bibinfo{volume}{D63}},
  \bibinfo{pages}{054008} (\bibinfo{year}{2001}), \eprint{hep-ph/0004173}.

\bibitem{pqcd3}
\bibinfo{author}{\bibfnamefont{H.~N.} \bibnamefont{Li}},
  \bibinfo{journal}{Phys. Rev.} \textbf{\bibinfo{volume}{D52}},
  \bibinfo{pages}{3958} (\bibinfo{year}{1995}), \eprint{hep-ph/9412340}.


\bibitem{lu1}
\bibinfo{author}{\bibfnamefont{C.-D.} \bibnamefont{Lu}} \bibnamefont{and}
  \bibinfo{author}{\bibfnamefont{K.}~\bibnamefont{Ukai}},
  \bibinfo{journal}{Eur. Phys. J.} \textbf{\bibinfo{volume}{C28}},
  \bibinfo{pages}{305} (\bibinfo{year}{2003}), \eprint{hep-ph/0210206}.

\bibitem{lu2}
\bibinfo{author}{\bibfnamefont{Y.}~\bibnamefont{Li}} \bibnamefont{and}
  \bibinfo{author}{\bibfnamefont{C.-D.} \bibnamefont{Lu}},
  \bibinfo{journal}{Commun. Theor. Phys.} \textbf{\bibinfo{volume}{44}},
  \bibinfo{pages}{659} (\bibinfo{year}{2005}), \eprint{hep-ph/0502038}.

\bibitem{aguilar1}
\bibinfo{author}{\bibfnamefont{A.~C.} \bibnamefont{Aguilar}} \bibnamefont{and}
  \bibinfo{author}{\bibfnamefont{A.~A.} \bibnamefont{Natale}},
  \bibinfo{journal}{JHEP} \textbf{\bibinfo{volume}{08}}, \bibinfo{pages}{057}
  (\bibinfo{year}{2004}), \eprint{hep-ph/0408254}.

\bibitem{aguilar2a}
\bibinfo{author}{\bibfnamefont{A.~C.} \bibnamefont{Aguilar}} \bibnamefont{and}
  \bibinfo{author}{\bibfnamefont{J.}~\bibnamefont{Papavassiliou}},
  \bibinfo{journal}{JHEP} \textbf{\bibinfo{volume}{12}}, \bibinfo{pages}{012}
  (\bibinfo{year}{2006}), \eprint{hep-ph/0610040}.

\bibitem{aguilar2b}
\bibinfo{author}{\bibfnamefont{A.~C.} \bibnamefont{Aguilar}} \bibnamefont{and}
  \bibinfo{author}{\bibfnamefont{J.}~\bibnamefont{Papavassiliou}}
  (\bibinfo{year}{2007}), \eprint{arXiv: 0708.\-4320 [hep-ph]}.

\bibitem{alkofer}
\bibinfo{author}{\bibfnamefont{C.~S.} \bibnamefont{Fischer}},
  \bibinfo{author}{\bibfnamefont{R.}~\bibnamefont{Alkofer}}, \bibnamefont{and}
  \bibinfo{author}{\bibfnamefont{H.}~\bibnamefont{Reinhardt}},
  \bibinfo{journal}{Phys. Rev.} \textbf{\bibinfo{volume}{D65}},
  \bibinfo{pages}{094008} (\bibinfo{year}{2002}), \eprint{hep-ph/0202195}.

\bibitem{alkofer2}
\bibinfo{author}{\bibfnamefont{R.}~\bibnamefont{Alkofer}} \bibnamefont{and}
  \bibinfo{author}{\bibfnamefont{L.}~\bibnamefont{von Smekal}},
  \bibinfo{journal}{Phys. Rept.} \textbf{\bibinfo{volume}{353}},
  \bibinfo{pages}{281} (\bibinfo{year}{2001}), \eprint{hep-ph/0007355}.

\bibitem{alkoferc}
\bibinfo{author}{\bibfnamefont{R.} \bibnamefont{Alkofer}},
  \bibinfo{author}{\bibfnamefont{C.~S.} \bibnamefont{Fischer}},
  \bibinfo{author}{\bibfnamefont{F.~J.} \bibnamefont{Llanes-Es\-tra\-da}},
  \bibnamefont{and}
  \bibinfo{author}{\bibfnamefont{K.}~\bibnamefont{Schwen\-zer}},
  \bibinfo{journal}{POS} \textbf{\bibinfo{volume}{LAT2007}},
  \bibinfo{pages}{286} (\bibinfo{year}{2007}), \eprint{arXiv:0710.1154
  [hep-ph]}.

\bibitem{ball}
\bibinfo{author}{\bibfnamefont{P.}~\bibnamefont{Ball}}, \bibinfo{journal}{JHEP}
  \textbf{\bibinfo{volume}{09}}, \bibinfo{pages}{005} (\bibinfo{year}{1998}),
  \eprint{hep-ph/9802394}.

\bibitem{belle}
\bibinfo{author}{\bibfnamefont{K.}~\bibnamefont{Abe}} \bibnamefont{et~al.}
  (\bibinfo{collaboration}{Belle}), \bibinfo{journal}{Phys. Rev. Lett.}
  \textbf{\bibinfo{volume}{98}}, \bibinfo{pages}{181804}
  (\bibinfo{year}{2007}), \eprint{hep-ex/0608049}.

\bibitem{pdg}
\bibinfo{author}{\bibfnamefont{W.~M.} \bibnamefont{Yao}} \bibnamefont{et~al.}
  (\bibinfo{collaboration}{Particle Data Group}), \bibinfo{journal}{J. Phys.}
  \textbf{\bibinfo{volume}{G33}}, \bibinfo{pages}{1} (\bibinfo{year}{2006}).

\bibitem{babar}
\bibinfo{author}{\bibfnamefont{B.}~\bibnamefont{Aubert}} \bibnamefont{et~al.}
  (\bibinfo{collaboration}{BABAR}) (\bibinfo{year}{2005}),
  \eprint{hep-ex/0508046}.

\bibitem{cdf1}
\bibinfo{author}{\bibfnamefont{A.}~\bibnamefont{Abulencia}}
  \bibnamefont{et~al.} (\bibinfo{collaboration}{CDF}), \bibinfo{journal}{Phys.
  Rev. Lett.} \textbf{\bibinfo{volume}{97}}, \bibinfo{pages}{211802}
  (\bibinfo{year}{2006}), \eprint{hep-ex/0607021}.

\bibitem{cdf2}
\bibinfo{author}{\bibfnamefont{M.}~\bibnamefont{Morello}}
  (\bibinfo{collaboration}{CDF}), \bibinfo{journal}{Nucl. Phys. Proc. Suppl.}
  \textbf{\bibinfo{volume}{170}}, \bibinfo{pages}{39} (\bibinfo{year}{2007}),
  \eprint{hep-ex/0612018}.

\bibitem{belle2}
\bibinfo{author}{\bibfnamefont{P.}~\bibnamefont{Krokovny}} \bibnamefont{et~al.}
  (\bibinfo{collaboration}{Belle}), \bibinfo{journal}{Phys. Rev. Lett.}
  \textbf{\bibinfo{volume}{89}}, \bibinfo{pages}{231804}
  (\bibinfo{year}{2002}), \eprint{hep-ex/0207077}.

\bibitem{babar2}
\bibinfo{author}{\bibfnamefont{B.}~\bibnamefont{Aubert}} \bibnamefont{et~al.}
  (\bibinfo{collaboration}{BABAR}), \bibinfo{journal}{Phys. Rev. Lett.}
  \textbf{\bibinfo{volume}{90}}, \bibinfo{pages}{181803}
  (\bibinfo{year}{2003}), \eprint{hep-ex/0211053}.

\bibitem{wilson}
\bibinfo{author}{\bibfnamefont{C.~D.} \bibnamefont{Lu}},
  \bibinfo{author}{\bibfnamefont{K.}~\bibnamefont{Ukai}}, \bibnamefont{and}
  \bibinfo{author}{\bibfnamefont{M.~Z.} \bibnamefont{Yang}},
  \bibinfo{journal}{Phys. Rev.} \textbf{\bibinfo{volume}{D63}},
  \bibinfo{pages}{074009} (\bibinfo{year}{2001}), \eprint{hep-ph/0004213}.

\bibitem{aguilar08}
\bibinfo{author}{\bibfnamefont{A.~C.} \bibnamefont{Aguilar}},
  \bibinfo{author}{\bibfnamefont{D.}~\bibnamefont{Binosi}}, \bibnamefont{and}
  \bibinfo{author}{\bibfnamefont{J.}~\bibnamefont{Papavassiliou}}
  (\bibinfo{year}{2008}), \eprint{arXiv:0802.1870 [hep-ph]}.

\end{thebibliography}

\end{document}